\documentclass[namedreferences]{SolarPhysics}
\usepackage[optionalrh]{spr-sola-addons} % For Solar Physics 
\usepackage{graphicx}        % For eps figures, newer & more powerfull
\usepackage{color}           % For color text: \color command
\usepackage{url}             % For breaking URLs easily trough lines
\begin{document}

\begin{article}

\begin{opening}

\title{Resonantly Damped Surface and Body MHD Waves in a Solar Coronal Slab with Oblique 
Propagation}

\author{I.~\surname{Arregui}$^{1}$\sep
        J.~\surname{Terradas}$^{1,2}$\sep
        R.~\surname{Oliver}$^{1}$\sep
        J.~L.~\surname{Ballester}$^{1}$       
       }
\runningauthor{Arregui {\it et al.}}
\runningtitle{Resonantly Damped Surface and Body MHD Waves}

   \institute{$^{1}$ Departament de F\'{\i}sica, Universitat de les Illes Balears, E-07122 
                     Palma de Mallorca, Spain
                     email: \url{inigo.arregui@uib.es}\\ 
              $^{2}$ Centrum  voor Plasma Astrofysica, KULeuven,
                     Celestijnenlaan 200B, 3001 Heverlee, Belgium
                     \\
             }

\begin{abstract}
The theory of magnetohydrodynamic (MHD) waves in solar coronal slabs in a zero-$\beta$ 
configuration and for parallel propagation of waves does not allow the existence of surface waves. 
When oblique propagation of perturbations is considered both surface and body  waves are 
able to propagate. When the perpendicular wave number is larger than a certain value, the body kink 
mode becomes a surface wave. In addition, a sausage surface mode is found below the internal cut-off 
frequency. When non-uniformity in the equilibrium is included, surface and body modes are damped due 
to resonant absorption. In this paper, first, a normal-mode analysis is performed and the period, 
the damping rate, and the spatial structure of eigenfunctions are obtained. Then, the time-dependent 
problem is solved, and the conditions under which one or the other type of mode is excited are 
investigated.
\end{abstract}
\end{opening}
%-------------------------------------------------

\section{Introduction}\label{introduction} 

Coronal seismology, such as first suggested by Uchida (1970) and Roberts, Edwin, and Benz (1984), 
aims to determine unknown physical parameters in the solar corona by the combination of
observed and theoretical properties of waves and oscillations. 
Flare-generated transverse coronal-loop oscillations have attracted particular interest, 
since their first unambiguous detection with high resolution instruments on-board spacecraft, 
such as SOHO and TRACE (Aschwanden {\it et al.} 1999; Nakariakov {\it et al.} 1999; Aschwanden {\it et al.} 2002; 
Schrijver, Aschwanden, and Title, 2002). The observed motions have been interpreted in terms of fast 
MHD kink modes of a cylindrical flux tube in their fundamental harmonic (Nakariakov {\it et al.} 1999), 
opening the way of coronal seismology. Recent examples of the application of coronal seismology can be 
found in Nakariakov and Ofman (2001); Goossens, Andries, and Aschwanden (2002); Aschwanden {et al.} (2003);
Andries, Arregui, and Goossens (2005); Verwichte, Foullon, and Nakariakov (2006c); Arregui {\it et al.} (2007b).

The basic theoretical framework for MHD wave propagation in structured media was developed 
well in advance of the observational evidence for oscillations.  For example, Edwin and 
Roberts (1982) studied wave propagation in coronal loops modelled as straight magnetic slabs. 
Straight cylindrical flux tube models were used by Spruit (1981); Edwin and Roberts (1983);  
Roberts (1983). More recent models have explored other effects, such as
the curvature of coronal loops (Van Doorsselaere {\it et al.} 2004; Brady and Arber,  2005; 
D\'{\i}az, 2006; D\'{\i}az, Zaqarashvili, and Roberts, 2006; Terradas, Oliver, and Ballester, 2006;
Verwichte, Foullon, and Nakariakov, 2006a, b);
the non-circularity of their cross-sections (Ruderman, 2003) or the influence of a longitudinally-varying 
density (Andries {\it et al.} 2005; Andries, Arregui, and Goossens 2005; Arregui {\it et al.} 2005; 
McEwan {\it et al.} 2006; Dymova and Ruderman 2006). Recent comprehensive reviews on these and 
other studies can be found in Nakariakov and Verwichte (2005) and Aschwanden (2006).

Among all these studies, surface waves have been the subject of considerable attention. 
Jain and Roberts (1996) studied the properties of magneto-acoustic surface waves at a single 
interface, for the case of non-parallel propagation, in the context of {\it f}-mode oscillations. 
In a different context Zhelyazkov, Murawski, and Goossens (1996) considered the propagation of 
magneto-acoustic surface waves with both non-parallel propagation and a sheared magnetic field. 
Similar studies of surface waves can be found in Roberts (1981a) and Miles and Roberts (1989).
By surface wave we mean a wave with an evanescent tail that propagates on a sharp (discontinuous) 
interface, although they can also propagate on a smoothly varying interface 
(see Lee and Roberts, 1986; Hollweg, 1990a, b, 1991; Goossens, 1991, for example). Surface waves are 
expected to occur in the solar atmosphere at places where physical parameters change abruptly. 
Their propagation is directional and guided by the interface and their energy is confined to 
within roughly a wavelength of the surface. Surface waves may also arise in more complex structures 
than a single interface, such as slabs and  flux tubes. However, the existence of surface waves 
in a zero-$\beta$ equilibrium requires the presence of propagation in the direction perpendicular
to the magnetic field (Roberts, 1991).

In this paper, damped coronal-loop oscillations (Aschwanden {\it et al.} 1999; Nakariakov {\it et al.} 1999) 
are considered. We adopt a line-tied slab model for a coronal loop and study the MHD 
wave properties, for oblique propagation of waves, by solving the normal-mode 
problem as well as the time-dependent problem.  
The non-uniformity of the equilibrium produces the resonant damping of these oscillations 
(Goossens, Andries, and Arregui, 2006). Although the model adopted in this work is an oversimplification, 
and is far from being an accurate representation of real coronal loops, the simplicity of the model 
allows a detailed study of the problem and a comparison with the results in a cylindrical loop model.

The layout of the paper is as follows. In Section~\ref{eq}, the equilibrium configuration and the 
basic MHD equations governing resonantly-coupled fast and Alfv\'en modes are presented. Next, in 
Section~\ref{normal}, the normal mode properties are described. In Section~\ref{timesect}, we analyse 
the temporal evolution of the perturbations after given initial disturbances. Finally, in 
Section~\ref{conclusions}, our conclusions are drawn.

\section{Equilibrium Configuration, Linear MHD Waves, and Numerical Method}\label{eq}

\begin{figure}[!t]
\begin{center}
\includegraphics[width=7cm,angle=90]{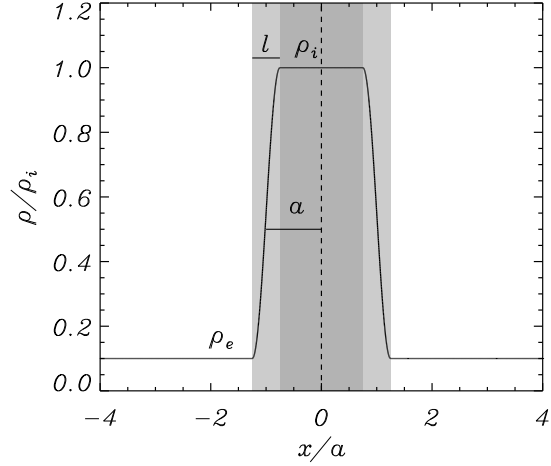}
\end{center} 
\caption{Schematic representation of the density enhancement (shaded region) of half-width 
$a$ located at $x=0$ used to model a coronal loop. The internal region with density $\rho_i$ 
is connected to the external medium, with density $\rho_e$, by transitional non-uniform 
layers (light-shaded regions) of thickness $l$. Non-uniform layers of thickness $l/a=0.5$ have 
been considered. }
         \label{model}
\end{figure}

We model the equilibrium magnetic and plasma configuration of a solar coronal loop by means
of a one-dimensional, line-tied, over-dense slab in Cartesian geometry. 
The magnetic field is straight and pointing in the $z$-direction, ${\bf B}=B\ \hat{\bf e}_z$. 
For applications to the solar corona it is a good approximation to consider that
the magnetic pressure dominates over the gas
pressure. This classic zero plasma-$\beta$ limit 
implies that the magnetic field is uniform and
that the density [$\rho(x)$] or Alfv\'en
speed [$v_{A}(x)$] profiles can be chosen arbitrarily. 
The coronal slab is then modelled
using a varying  equilibrium density
profile in the $x$-direction (see Figure~\ref{model}), by means of a density enhancement
of half-width $a$ centred about $x=0$.  The
density inside the slab is constant [$\rho_i$] and
connected to the constant coronal environment, with density $\rho_e$, by transitional
non-uniform layers of thickness $l$. The explicit expression for the
equilibrium density considered is 

\begin{equation}
\rho(x)=\left\{\begin{array}{ll}
\rho_i \hspace{2.8cm} \textrm{$0\leq x \leq a-\frac{l}{2}$},\\\\
f(x) \hspace{1.55cm} \textrm{$a-\frac{l}{2}\leq x \leq a+\frac{l}{2}$},\\\\
\rho_e \hspace{3.5cm} \textrm{$x\geq a+\frac{l}{2}$},\\\\
\end{array}\right\}\label{profile2slabs}
\end{equation}

\noindent
with $\rho(-x)=\rho(x)$.
The density profiles at the non-uniform transitional layers
have been chosen following Ruderman and Roberts (2002); Van Doorsselaere {\it et al.} (2004)
and are given by 

\begin{eqnarray}
f(x)&=&\frac{\rho_i}{2}\left[\left(1+\frac{\rho_e}{\rho_i}\right)-\left(1-\frac{\rho_e}{\rho_i}
\right)\sin{\frac{\pi\left(x-a\right)}{l}}\right].\\\nonumber
\end{eqnarray}

\noindent
We remark that the particular form of the density profile at the non-uniform layers has no 
importance by itself for the results obtained.

In order to study small-amplitude oscillations of the previous equilibrium, the linear 
resistive MHD equations with constant magnetic diffusivity [$\eta$] are considered. 
The inclusion of dissipative terms in the MHD equations is needed for the computation of 
resonantly-damped eigenmodes. Resistive dissipation is preferred over viscous dissipation 
for the simplicity with which it enters into the equations. Both dissipation mechanisms 
affect the thickness of the dissipation layers and the behaviour of the solutions in those layers, 
but not the differences in the solutions across the layers, which in turn determine the damping rate of 
the eigenmodes (Sakurai, Goossens, and Hollweg 1991).
As the equilibrium configuration only depends on the $x$-coordinate, a spatial dependence 
of the form $\exp^{-i\left(k_y y+k_z z\right)}$ is assumed for all perturbed quantities, 
with $k_y$ and $k_z$ being the perpendicular and parallel wave numbers. The photospheric line-tying effect 
is then included by selecting the appropriate parallel wavenumber. This leads to the following set of 
differential equations for the two components of the velocity perturbation [$v_x$ and $v_y$] and 
the three components of the perturbed magnetic field [$b_{1x}$, $b_{1y}$, and  $b_{1z}$] 

%%%%%%%%%%%%%%%%%%%%%%%%%%%%%%%%%%%%%%%%%%%%%%%%%%%%%%%%%%%%

%%%%%%%%%%%%%%%%%%%%%%%%%%%%%%%%%%%%%%%%%%%%%%%%%%%%%%%%%%%%

\begin{eqnarray}
\frac{\partial v_x}{\partial t} &=& \frac{B}{\mu\rho}\left(-i k_z b_{1x}- 
\frac{\partial b_{1z}}{\partial x}\right), \label{first}\\
\frac{\partial v_y}{\partial t} &=& \frac{B}{\mu\rho}\left(-i k_z b_{1y}+
i k_y b_{1z}\right),\\
\frac{\partial b_{1x}}{\partial t}&=&-i B k_z v_x +
\eta\left[\frac{\partial^2 b_{1x}}{\partial x^2}-\left(k^2_y+k^2_z\right) b_{1x}\right],\\
\frac{\partial b_{1y}}{\partial t}&=&-i B k_z v_y +\eta\left[\frac{\partial^2 b_{1y}}
{\partial x^2}-\left(k^2_y+k^2_z\right) b_{1y}\right],\\
\frac{\partial b_{1z}}{\partial t}&=&-B\left(\frac{\partial v_x}{\partial x}-i k_y v_y\right)+
\eta\left[\frac{\partial^2 b_{1z}}{\partial x^2}-\left(k^2_y+k^2_z\right) b_{1z}\right],\label{last}
\end{eqnarray}

\noindent
where $\mu$ is the permeability of free space. 
When $k_y\neq0$ and $l/a\neq 0$, Equations~(\ref{first})--(\ref{last}), describe the oscillatory properties of 
coupled fast and Alfv\'en modes. As the plasma-$\beta$ = $0$, the slow mode is absent and there are 
no motions parallel to the equilibrium magnetic field, $v_z=0$. In this paper, solutions to 
these equations are obtained by performing two kinds of analysis. First the normal modes of 
oscillation of the system are studied and then the time-dependent evolution of the slab after an 
initial disturbance is investigated. When only  oblique propagation is considered ($k_y\neq0$), 
the normal-mode solutions can be obtained by solving an analytical dispersion relation. If, in addition, 
transitional non-uniform layers are included ($l\neq0$), the solutions to these equations are, 
in general, difficult to obtain. For this reason, numerical approximations are obtained  using  
PDE2D (Sewell, 2005), a general-purpose partial differential equation solver. 
As for the boundary conditions, in both the normal mode analysis 
as well as in time-dependent simulations, we impose the vanishing of the perturbed velocity 
far away from the loop, hence ${\bf v}\rightarrow{ 0}$ as $x\rightarrow\pm\infty$.

\section{Normal Mode Analysis}\label{normal}

We first consider the normal modes of oscillation of our equilibrium configuration. A  
temporal dependence of the form $e^{i\omega t}$ is assumed for all perturbed quantities in 
Equations~(\ref{first})--(\ref{last}), with  $\omega=\omega_{R}+i \omega_{I}$ being  the complex 
frequency. These equations form an eigenvalue problem. The solutions for the simplest case, 
with $k_y=0$ and $l=0$, are well-known and described by Edwin and Roberts (1982). 
They can be classified, according to the parity of their eigenfunctions about $x=0$, as fast kink and 
sausage modes. The corresponding dispersion relations are

\begin{equation}\label{kink}
\tanh \kappa_i a=-\frac{\kappa_e}{\kappa_i},
\end{equation}

\noindent
for kink modes ($v_x$ even about $x=0$) and

\begin{equation}\label{sausage}
\coth \kappa_i a=-\frac{\kappa_e}{\kappa_i},
\end{equation}

\noindent
for sausage modes ($v_x$ odd  about $x=0$), where

\begin{equation}
\kappa^2_e=k^2_z-\frac{\omega^2}{v^2_{Ae}} \mbox{\hspace{1cm}} \mbox{and} 
\mbox{\hspace{1cm}} \kappa^2_i=k^2_z-\frac{\omega^2}{v^2_{Ai}}.
\end{equation}

\noindent
Here $v_{Ai,e}=B\sqrt{1/\mu\rho_{i,e}}$  are the internal and external Alfv\'en speeds.
Trapped waves have real frequencies whereas modes with complex frequency represent leaky waves. 
For the modes with the lowest number of internal nodes, the body sausage mode is leaky in 
the long wavelength limit, $k_z a$$\ll$$1$, while the kink mode is trapped for all wavelengths. 
All other branches, kink and sausage alike, are leaky in the long-wavelength limit.

The propagation of fast magneto-acoustic waves in a plasma density inhomogeneity with  smooth
transversal density profiles was considered by Edwin and Roberts (1988) and Nakariakov 
and Roberts (1995). The latter found that the simple slab profile is a good general guide to the 
behaviour in a smooth, sharply-structured, profile.
Here, a similar smooth transverse density profile  is
considered and oblique propagation of perturbations included.  
We will restrict our analysis to trapped modes, whereas for an analysis of the 
leaky solutions with $k_y=0$ the reader is referred to Terradas, Oliver, and Ballester (2005).

\subsection{Undamped Surface and Body Waves for Oblique Propagation}\label{undamped}

We first consider the changes of the oscillatory properties of fast modes due to the inclusion of oblique
propagation of perturbations.
%When oblique propagation of waves is considered the resulting solutions no longer represent 
%pure fast magneto-acoustic waves, but are coupled to Alfv\'en waves. 
%For the fundamental kink mode of oscillation in a cylinder this is always true since for this mode, 
%the azimuthal wave number is $m=1$ and there is always an dependence of perturbations on this 
%direction. 
If the slab is connected to the external medium by discontinuities ($l=0$) 
Equations~(\ref{first})--(\ref{last}) can be combined to give a single ordinary differential 
equation for the transverse component of the perturbed velocity of the form

\begin{equation}
\frac{{\mathrm d}\  }{{\mathrm d} x}\left[\frac{\rho v^2_{A}\left(k^2_z v^2_{A}-\omega^2\right)}
{(k^2_y+k^2_z) v^2_{A}-\omega^2} \frac{{\mathrm d}v_x  }{{\mathrm d} x} \right]-\rho\left(k^2_z v^2_{A}-\omega^2\right)v_x=0.
\end{equation}

\noindent
Solutions to this equation can readily be obtained,  for $\rho$ and $v_{A}$ constant, by following 
the usual procedure of matching different solutions in the internal and external regions 
(Edwin and Roberts, 1982) and demanding the evanescence of perturbations far away 
from the slab. This leads to the following dispersion relations

\begin{equation}\label{kinkky}
\tanh m_i a=-\frac{\kappa^2_e}{\kappa^2_i}\frac{m_i}{m_e},
\end{equation}

\noindent
for kink modes and

\begin{equation}\label{sausageky}
\coth m_i a=-\frac{\kappa^2_e}{\kappa^2_i}\frac{m_i}{m_e},
\end{equation}

\noindent
for sausage modes, where 

\begin{equation}\label{ms}
m^2_e=\left(k^2_y+k^2_z-\frac{\omega^2}{v^2_{Ae}}\right) \mbox{\hspace{1cm}} 
\mbox{and} \mbox{\hspace{1cm}} m^2_i=\left(k^2_y+k^2_z-\frac{\omega^2}{v^2_{Ai}}\right).
\end{equation}

\noindent
Equations~(\ref{kinkky}) and (\ref{sausageky}) reduce  to Equations~(\ref{kink}) and 
(\ref{sausage}) when $k_y$ = $0$, as expected. An interesting solution  of the dispersion relations 
can be obtained in the limit of quasi-perpendicular propagation, $k_y$$\gg$$k_z$, and additionally 
assuming that $k^2_y$$\gg$$\left(k^2_z-\omega^2/v^2_{Ai}\right)$. Then, the left hand-side of the 
dispersion relations given by Equations~(\ref{kinkky}) and (\ref{sausageky}) can be approximated to

\begin{equation}
\lbrace^{\tanh}_{\coth}\rbrace m_i a\simeq 1.
\end{equation}

\noindent
Let us define $\omega_{k}$ as the frequency in this limit. Then, 
the following expression for the parallel phase speed is obtained

\begin{equation}\label{kinkspeed}
\frac{\omega_k}{k_z}\simeq \sqrt{\frac{\rho_i v^2_{Ai}+\rho_e v^2_{Ae}}{\rho_i+\rho_e}}\equiv c_k,
\end{equation}

\noindent
where $c_k$ is the kink speed. This quantity arises also in the description 
of the kink mode of oscillation of a magnetic flux tube ({\it e.g.} Spruit, 1981; 
Ryutova, 1990; Roberts, 1991) and in the propagation of surface waves in the incompressible 
limit (Roberts, 1981b). 

\begin{figure*}[!t]
 
\includegraphics[width=10.0cm,angle=90]{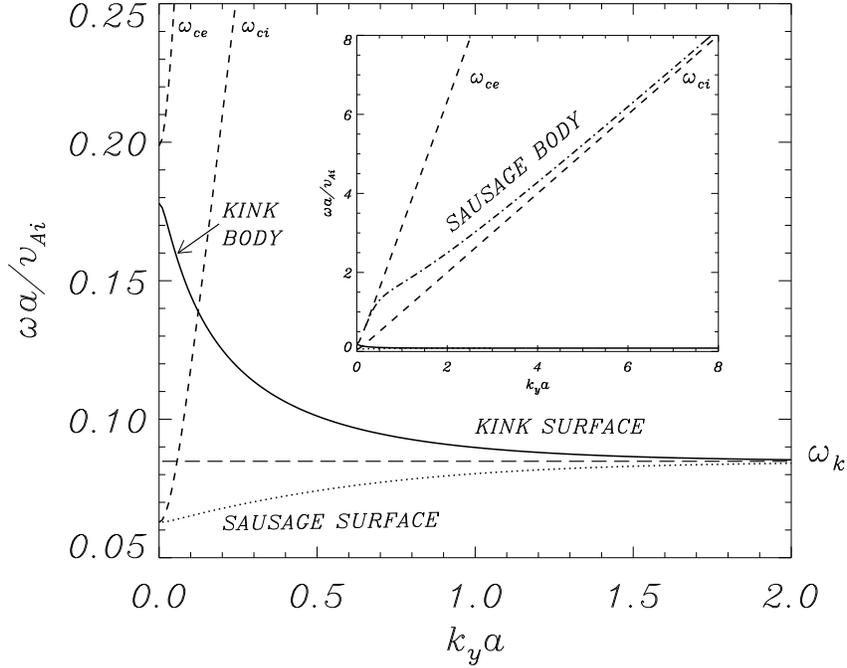}

  \caption{Dispersion diagram calculated from the numerical solution of the dispersion relations, 
Equations~(\ref{kinkky}) and (\ref{sausageky}), for a slab model of coronal loop with 
$k_z a=\pi/50$ and density contrast $\rho_i/\rho_e=10$. The frequency of the different solutions is
displayed as a function of the perpendicular wave number: kink mode (solid), sausage surface
mode (dotted) and sausage body (dash-dotted). The horizontal long-dashed line indicates the 
frequency $\omega_k$, given by Equation~(\ref{kinkspeed}), to which  both kink and sausage 
surface waves approach for large $k_y$. The inset plot displays a broader view of the 
solutions in order to show the behaviour 
of the sausage body solution. The dashed lines represent the internal and external cut-off 
frequencies, $w_{ci}=\sqrt{k^2_y+k^2_z}v_{Ai}$ and $\omega_{ce}=\sqrt{k^2_y+k^2_z}v_{Ae}$ respectively. 
For large $k_y$, the sausage body mode approaches asymptotically $\omega_{ci}$ never crossing below 
its value and the kink and sausage surface solutions go to the kink frequency given by 
Equation~(\ref{kinkspeed}).}
         \label{disperplots}
\end{figure*}

Solutions to Equations~(\ref{kinkky}) and (\ref{sausageky}) can be found by means of a 
simple numerical program. In our numerical calculations, we consider fixed values for the 
density contrast, $\rho_i/\rho_e=10$, and for the parallel wavenumber, $k_za=\pi/50$.  
For the observed transverse kink-mode oscillations, with a wavelength double the length of the loop, 
this corresponds to a ratio of length to width of $L/2a=25$, which is a typical value for observed 
oscillating coronal loops. Figure~\ref{disperplots} displays some solutions as a 
function of the perpendicular wave number. The dispersion relations have two frequencies at which 
the characteristics of wave 
motions vary. Modes with a frequency above the external cut-off frequency 
$\left[\omega_{ce}=v_{Ae}\sqrt{k^2_y+k^2_z}\right]$ are leaky. The internal cut-off frequency
$\left[\omega_{ci}=v_{Ai}\sqrt{k^2_y+k^2_z}\right]$ also plays an important role, since depending on whether 
the frequency of the eigenmode is above or below this cut-off we have a body-like or a 
surface-like mode.

%%%%%%%%%%%%%%%%%%%%%%%%%%%%%%%%%%%%%%%%%%%%%%%%%%%%%%%%%%%%%%%%%%%%%%%%%%%%%%%%%%%%%%%%%%%%%%%%%%%%%%
%%%%%%%%%%%%%%%%%%%%%%%%%%%%%%%%%%%%%%%%%%%%%%%%%%%%%%%%%%%%

%%%%%%%%%%%%%%%%%%%%%%%%%%%%%%%%%%%%%%%%%%%%%%%%%%%%%%%%%%%%%%%%%%%%%%%%%%%%%%%%%%%%%%%%%%%%%%%%
\begin{figure*}[!t]
 \hspace{-0.6cm}
\vbox{\hbox{
\includegraphics[width=5.cm,angle=90]{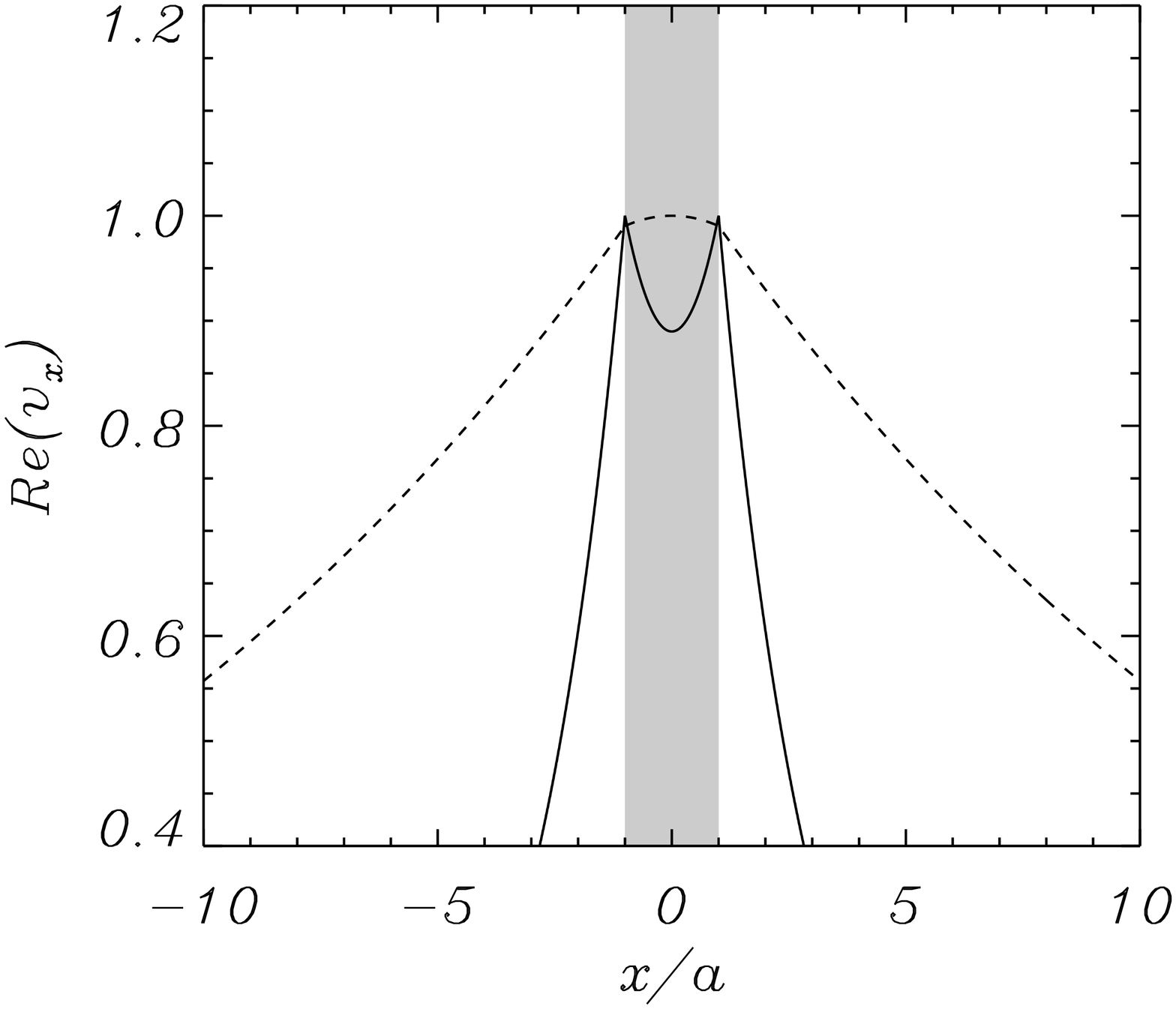}
  \includegraphics[width=5.cm,angle=90]{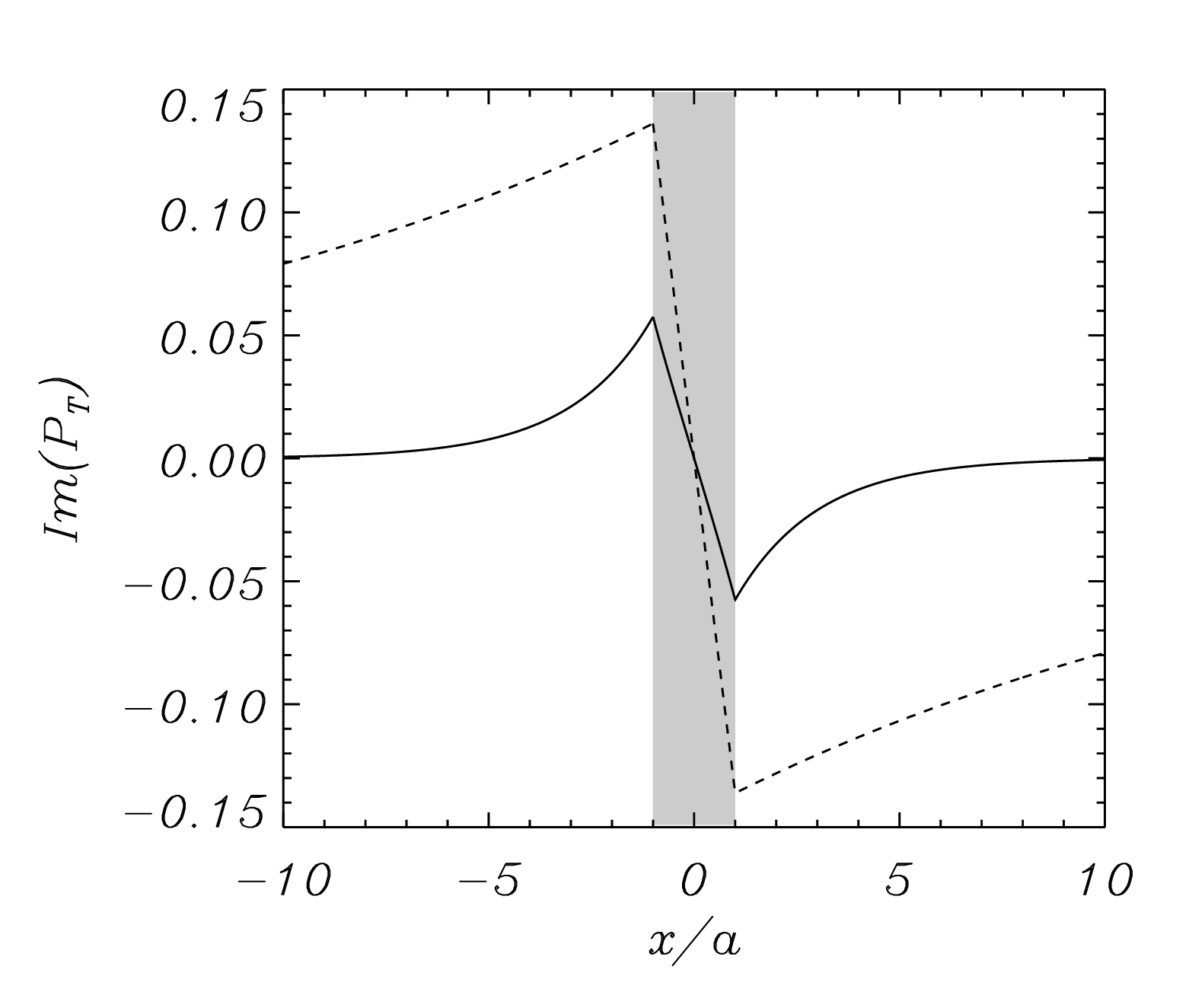} }
\hbox{
   \includegraphics[width=5.cm,angle=90]{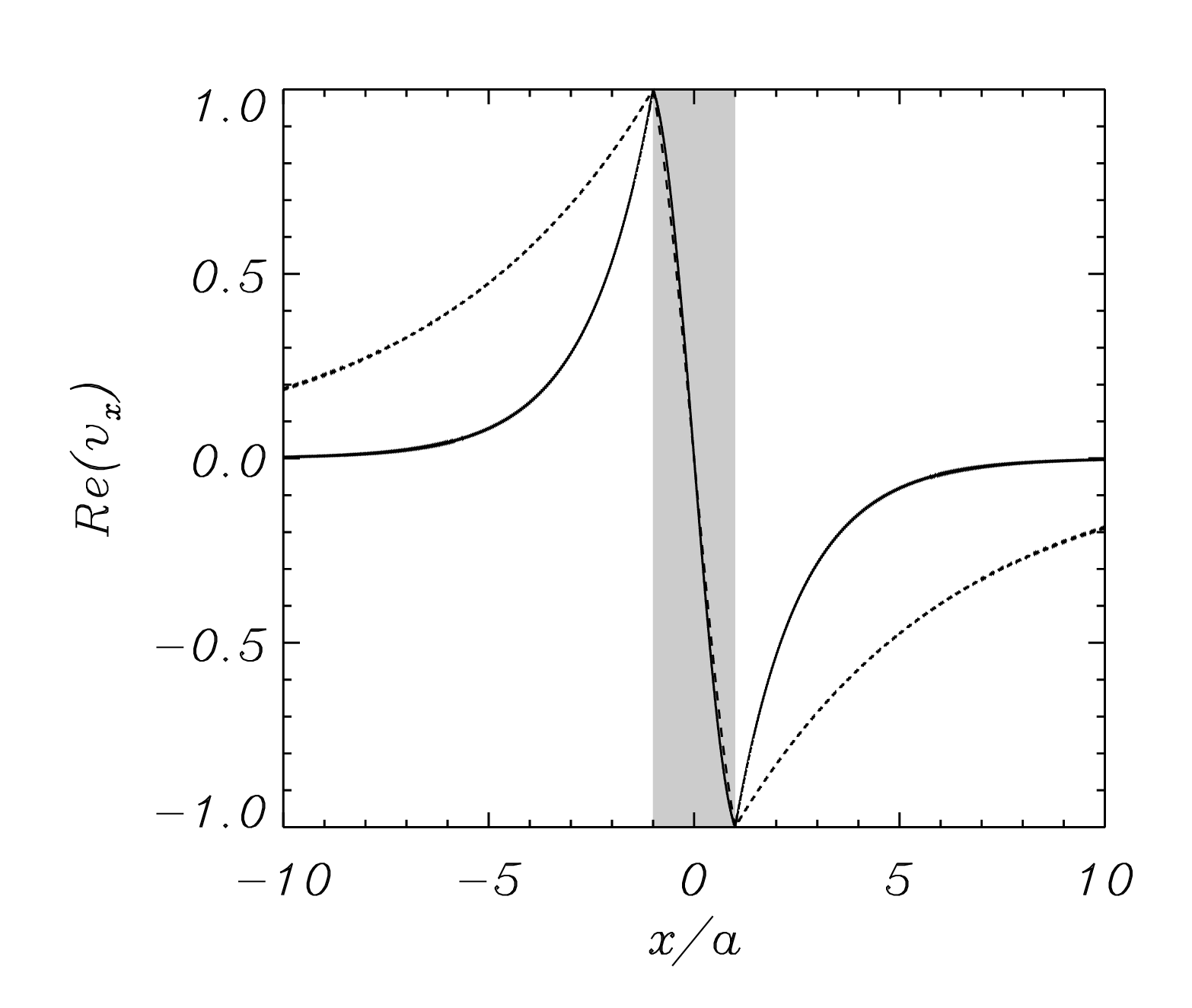}
    \includegraphics[width=5.cm,angle=90]{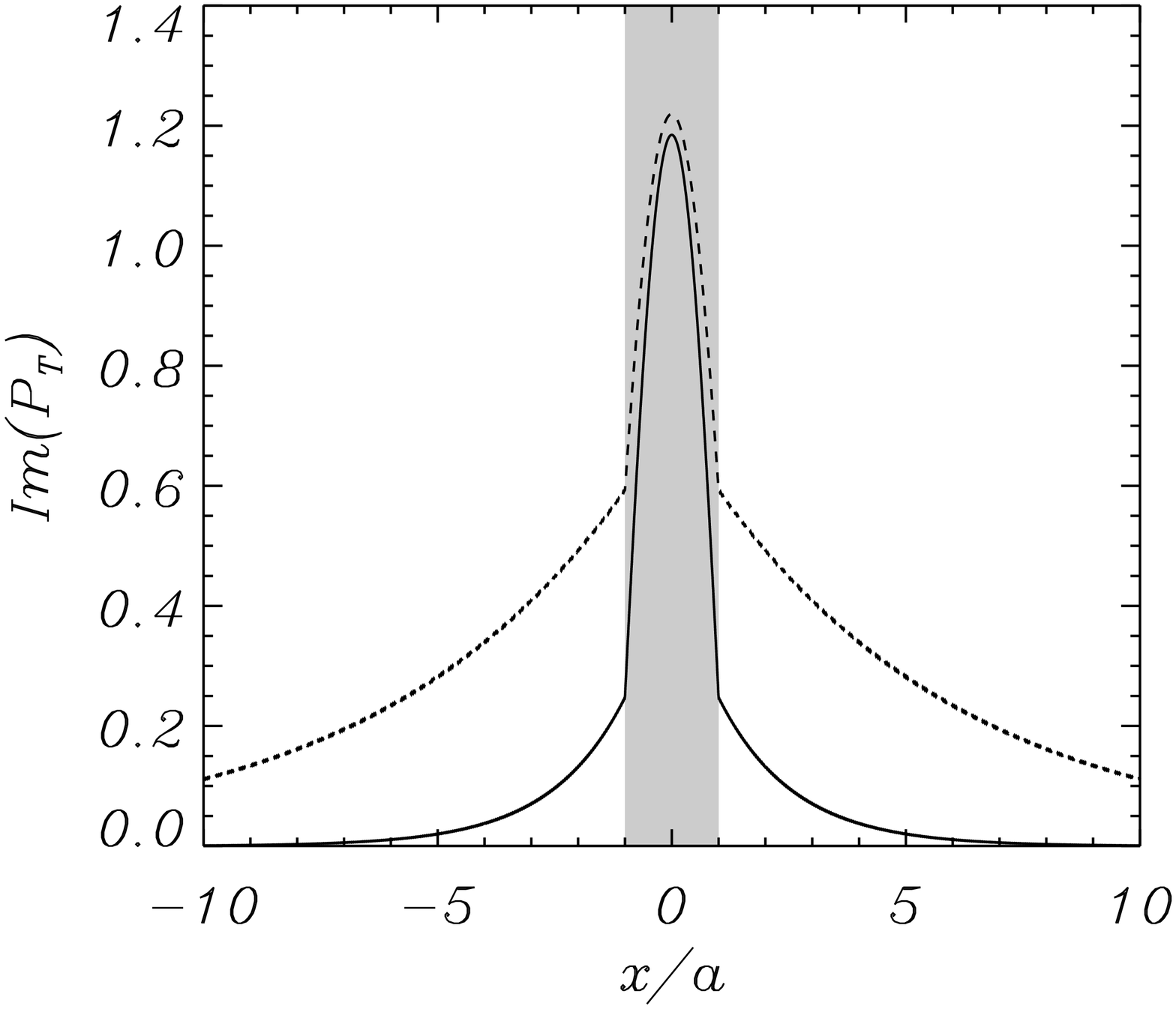}}
\hbox{
   \includegraphics[width=5.cm,angle=90]{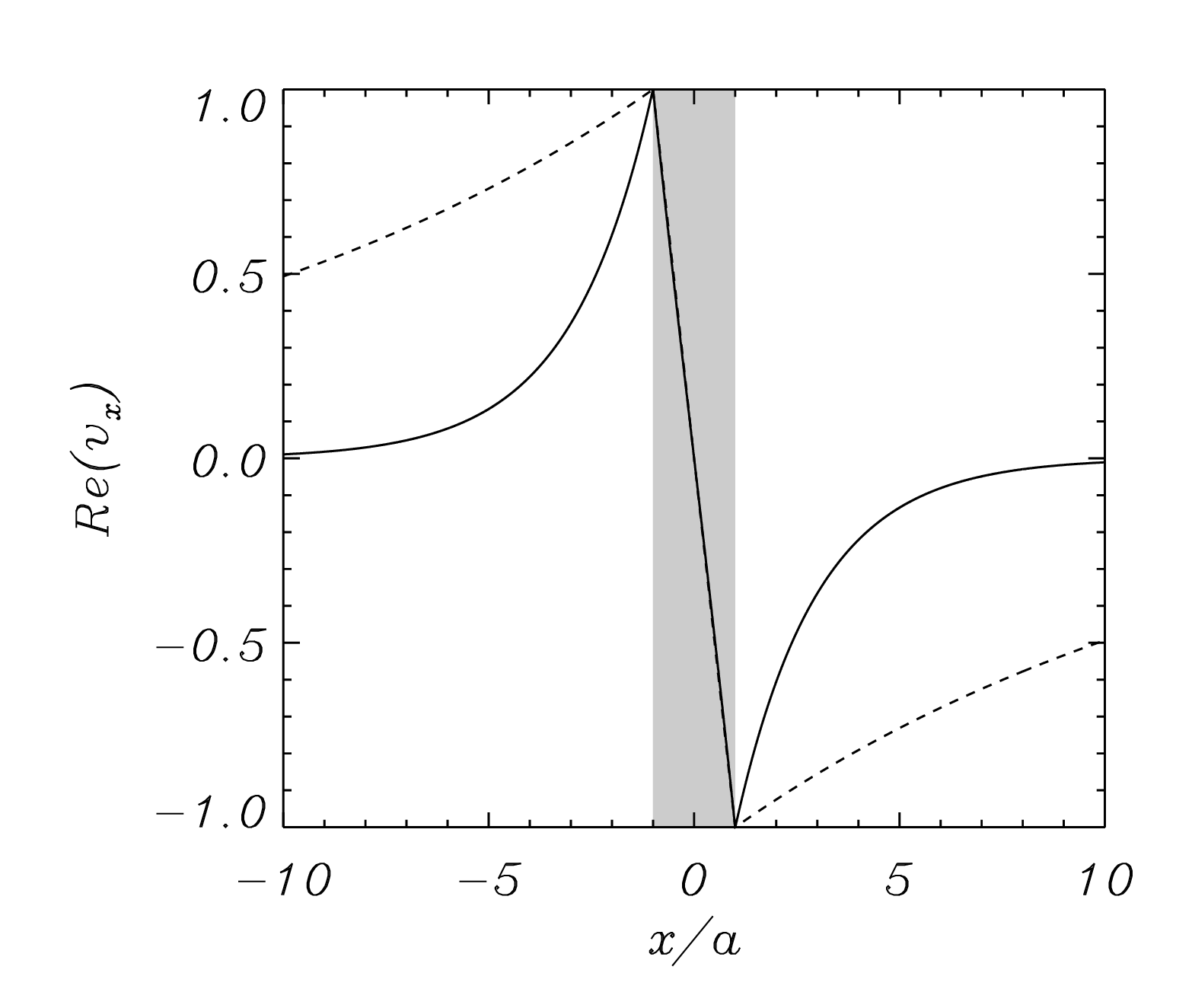}
    \includegraphics[width=5.cm,angle=90]{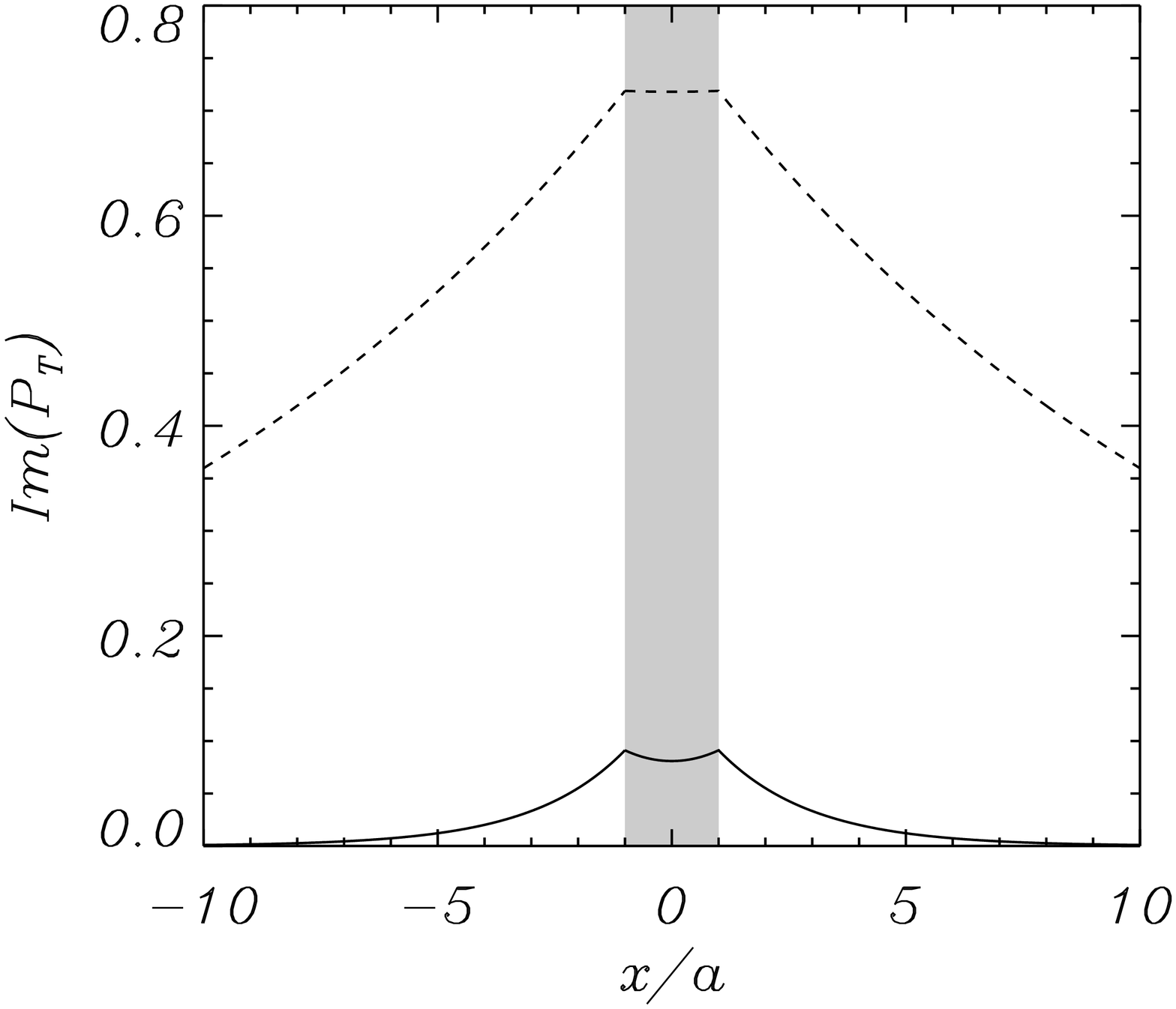}}}
\caption{Real part of the transverse velocity component [$v_x$] and imaginary part of the perturbed 
total pressure [$P_T$] for the fundamental kink mode ({\em top panels}), the sausage body mode 
({\em middle panels}) and the sausage surface mode ({\em bottom panels}) for an equilibrium 
configuration with $k_z a=\pi/50$, $\rho_i/\rho_e=10$ and two values of the perpendicular wave number: 
$k_y a=0.0$ (dotted lines) and $k_ya=0.5$ (solid lines) for the top and bottom panels and $k_y a=0.2$ 
(dotted lines) and $k_ya=0.5$ (solid lines) for the middle panel. Note that the character of the 
eigenfunction for the kink mode changes from body-like to surface-like and that for $k_y a=0.5$ the 
sausage body and surface mode solutions have an almost identical $v_x$. In all figures, 
the shaded region represents 
the density enhancement.}
         \label{eigen}
\end{figure*}

%%%%%%%%%%%%%%%%%%%%%%%%%%%%%%%%%%%%%%%%%%%%%%%%%%%%%%%%%%%%%%%%%%%%%%%%%%%%%%%%%%%%%%%%%%%%%%%%%%%%%%
The kink-mode solution has a decreasing frequency with respect to
$k_y$. For small values of $k_y$ its frequency is above the internal cut-off frequency 
($m^2_i<0$, in Equation~(\ref{kinkky})) and the mode is a body wave.
Beyond a given value of $k_y$ ($k_ya\sim 0.15$ in Figure~\ref{disperplots}), however, the frequency crosses  the 
internal cut-off frequency ($m^2_i>0$) and the mode becomes a surface
solution. In addition, as soon as $k_y\neq0$, there is also a  solution with frequency near 
the internal cut-off frequency. This solution is always below the
internal cut-off frequency and, therefore, corresponds to a surface wave. 
For this solution $v_x$ has odd parity with respect to $x=0$ and so is a sausage surface wave. 
For increasing $k_y$ the sausage surface mode 
approaches the kink mode solution and for large $k_y$ both
asymptotically tend to the frequency given by Equation~(\ref{kinkspeed}).
Therefore, for almost perpendicular propagation there are two surface modes corresponding to the 
two parities allowed by the symmetry of the problem. Contrary to the case $k_y=0$, the mode with
lowest frequency is sausage and not kink. The inset plot in Figure~\ref{disperplots} displays a 
broader view of the solutions. The sausage body solution is leaky for small values of $k_y$, but
it becomes non-leaky and its frequency is that of the external cut-off frequency near $k_ya=0.2$. 
Then, for increasing $k_y$ its frequency increases. In contrast to the kink-mode solution the 
frequency of the sausage body mode never crosses below the internal cut-off frequency, but approaches 
asymptotically its value for increasing $k_y$. The reason for this behaviour is that there is 
already a solution with odd parity with respect to $x=0$ below the internal cut-off frequency.

We obtain further information on the properties of the normal modes from the spatial distribution 
of the eigenfunctions. Figure~\ref{eigen} shows the perturbed transverse velocity [$v_x$] and the 
perturbed magnetic pressure [$P_T$ obtained from $b_{1z}$], for two different 
values of the perpendicular wave number and for the fundamental kink eigenmode, the sausage body mode 
and the sausage surface mode. Significant changes of the properties of eigenfunctions are 
clearly visible as the value of the perpendicular wave number is increased. For $k_y=0$, the kink 
mode has its  characteristic behaviour with a maximum of $v_x$ at the internal part of the slab, while  
$v_x$ decreases exponentially for increasing distance from the loop. On the other hand, the total 
pressure perturbation for this mode has the usual antisymmetric profile with a zero value at the 
centre of the slab and maxima at its edges. When oblique propagation is included ($k_y\neq0$), 
there is an increased confinement of the eigenfunctions, compared with non-oblique propagation. 
This result, also found by D\'{\i}az, Oliver, and Ballester (2003) in the context of prominence 
fibril oscillations, can  be seen in terms of a sharper drop-off rate of the mode in the external 
medium. The solutions for $k_ya\neq0$ over-plotted in Figure~\ref{eigen} correspond to a value for 
which the  kink solution is already below the internal cut-off frequency and show significant changes 
with respect to the $k_y=0$ case. The kink mode becomes a surface-like mode and motions are more 
confined to the edges of the slab. Further increase of $k_y$ produces a more marked confinement and a 
lower amplitude inside the slab. The middle and bottom panels of Figure~\ref{eigen} show the 
eigenfunctions for the body and surface sausage modes. The most important property is 
the similarity in the transverse component of the velocity for both solutions when $k_y$ is 
sufficiently large and also that the motions are very confined to the edges of the slab. 
The magnetic-pressure perturbation distribution peaks at the centre of the slab for the body 
solution, while it is maximum at the edges in the case of the surface solution. 
Finally, oblique propagation of waves affects the magnitude of the total pressure 
perturbation of the surface sausage mode, but not very much that of the body sausage mode 
inside the density enhancement.

\subsection{Damping of Oscillations by Resonant Absorption}\label{damped}

In this section the most general equilibrium configuration considered in this work is taken. 
The uniform internal density of the slab is now connected to the external (coronal) medium 
by means of non-uniform transitional layers. This produces damping by  resonant coupling of fast 
modes to Alfv\'en waves (Hollweg and Yang, 1988; Goossens, Andries, and Arregui, 2006).
In order to solve Equations~(\ref{first})--(\ref{last}), in this case we have to resort 
to numerical techniques and the numerical code PDE2D is used. 
The code uses finite elements and allows the use of non-uniformly distributed grids, which is needed 
in order to better resolve the large gradients that arise in the vicinity of the resonant layers 
and has been used successfully, in a similar problem, by Terradas, Oliver, and Ballester (2006). 
For the normal-mode analysis, a small but finite value of resistivity has to be provided when 
computing the damping of oscillations. Resistivity has to be small enough for the imaginary part 
of the frequency to be independent of resistivity  (Poedts and Kerner, 1991). This condition has 
been checked to a high accuracy for the solutions presented in this paper and a value for the 
magnetic Reynolds number,  $R_{m}/v_{Ai}a$,  between $10^6$ and $10^8$ has been sufficient.

%%%%%%%%%%%%%%%%%%%%%%%%%%%%%%%%%%%%%%%%%%%%%%%%%%%%%%%%%%%%%%%%%%%%%%%%%%%%%%%%%%%%%%%%%%%%%%
\begin{figure*}[!t]
 \begin{center}
\includegraphics[width=10.0cm,angle=90]{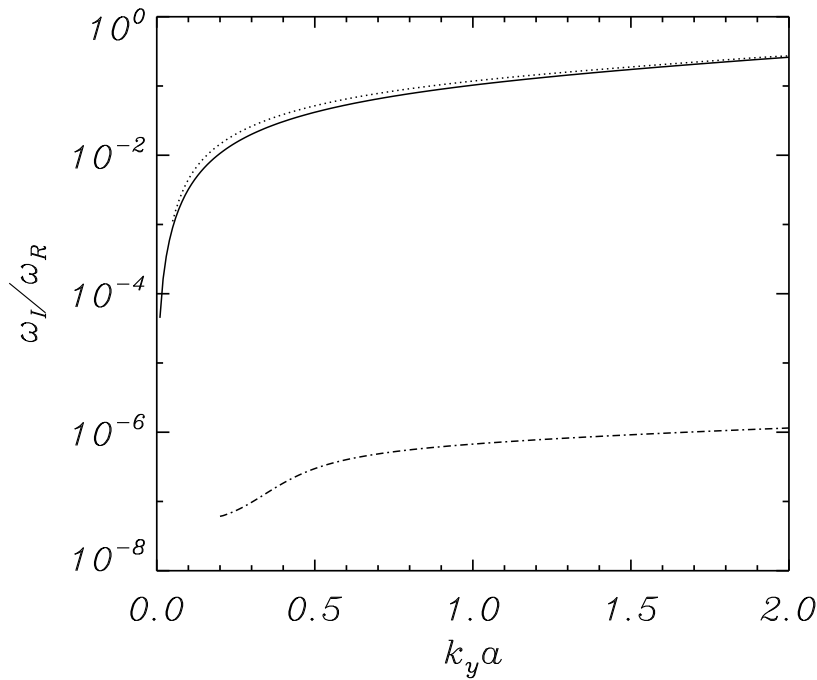}
\end{center}    
 \caption{Damping rate for the resonantly-damped fast kink surface mode (solid line), sausage 
surface mode (dotted line) and sausage body mode (dash-dotted line) as a function of the 
perpendicular wave number in a coronal slab model with $k_za=\pi/50$, $\rho_i/\rho_e=10$ and 
non-uniform transitional layers with thickness  $l/a=0.5$. A value for the magnetic Reynolds number 
of $R_{m}/v_{Ai}a=10^{7}$ has been used in the computation of the solutions in a non-uniform grid 
with $N_x=10\ 000$ points in the range 
$-50$ $\le x/a \le$ $50$.}
         \label{damping}
\end{figure*}
%%%%%%%%%%%%%%%%%%%%%%%%%%%%%%%%%%%%%%%%%%%%%%%%%%%%%%%%%%%%%%%%%%%%%%%%%%%%%%%%%%%%%%%%%%%%%%%%%%%
%%%%%%%%%%%%%%%%%%%%%%%%%%%%%%%%%%%%%%%%%%%%%%%%%%%%%%%%%%%%%%%%%%%%%%%%%%%%%%%%%%%%%%%%%%%%%%%%%%%
\begin{figure*}[!t]
 \hspace{-0.6cm}
\vbox{\hbox{
\includegraphics[width=5.cm,angle=90]{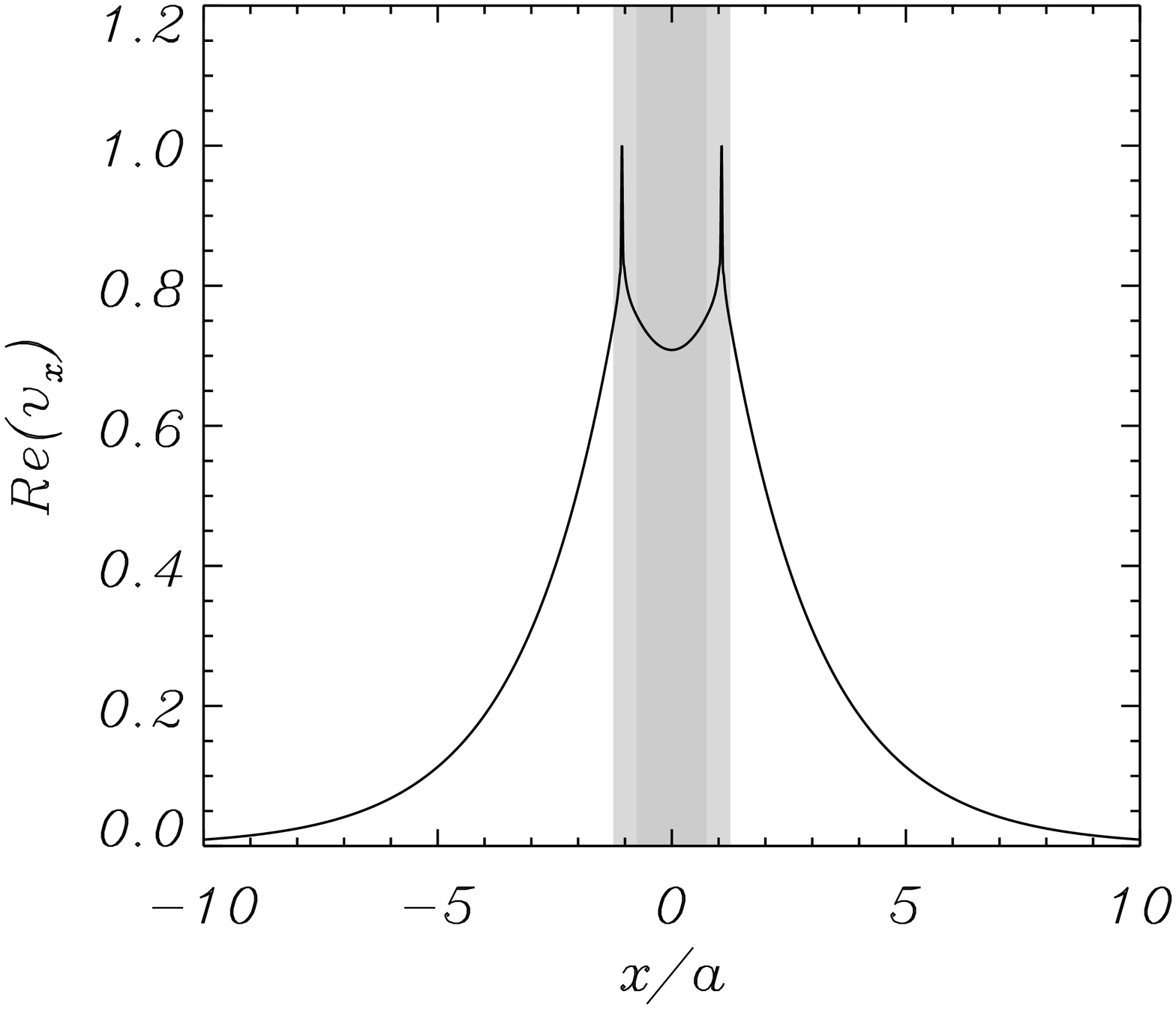}
  \includegraphics[width=5.cm,angle=90]{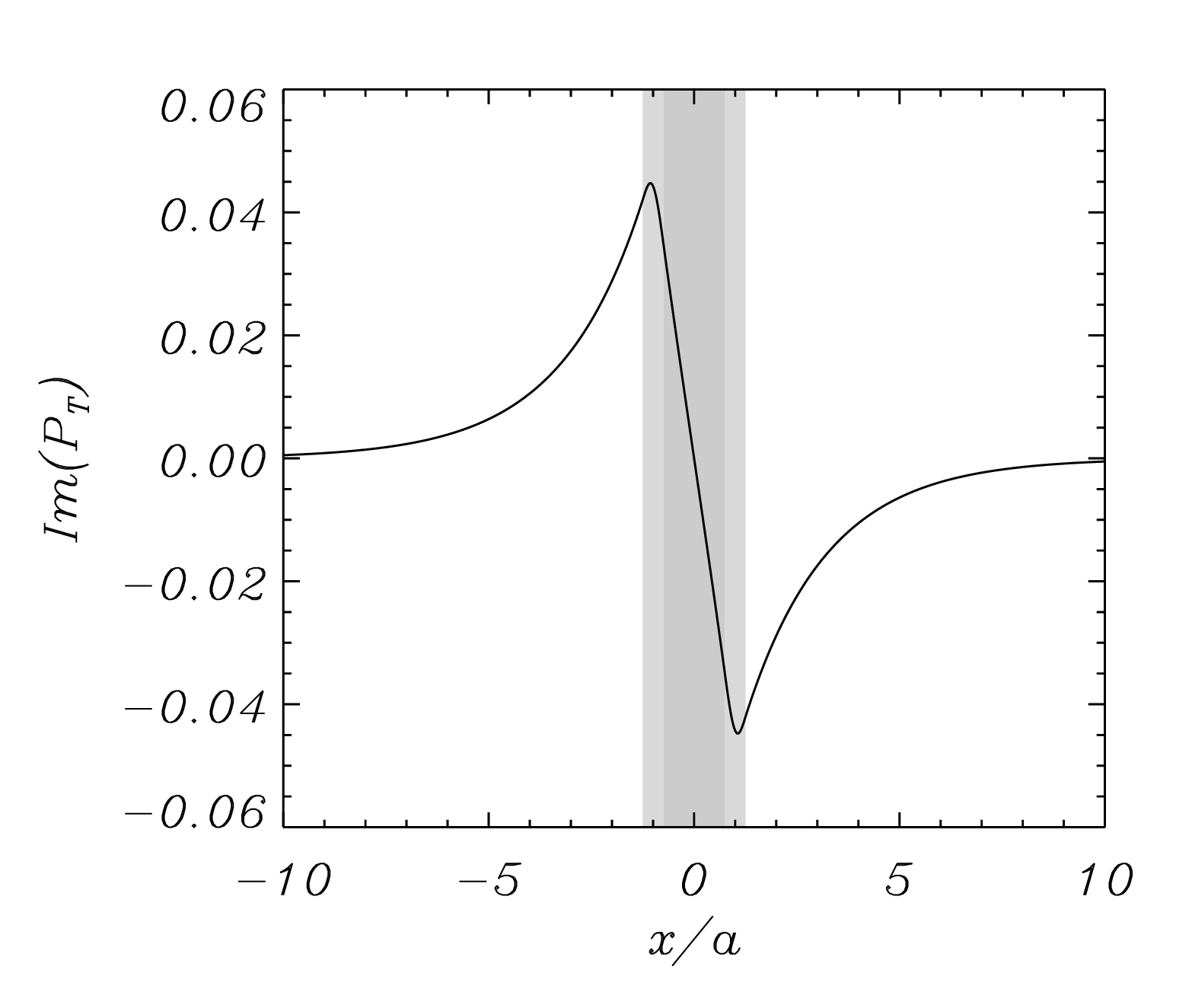} }
\hbox{
   \includegraphics[width=5.cm,angle=90]{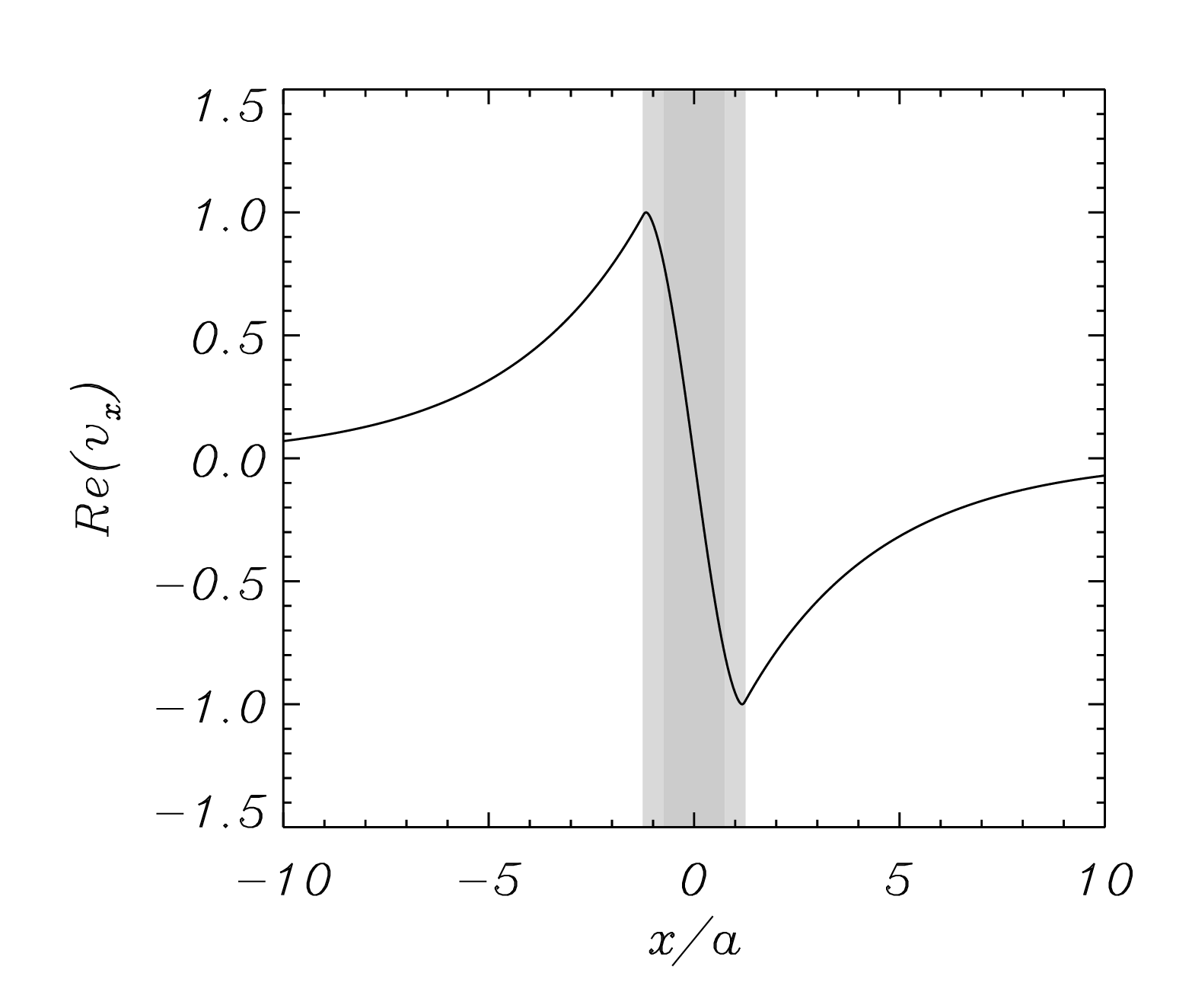}
    \includegraphics[width=5.cm,angle=90]{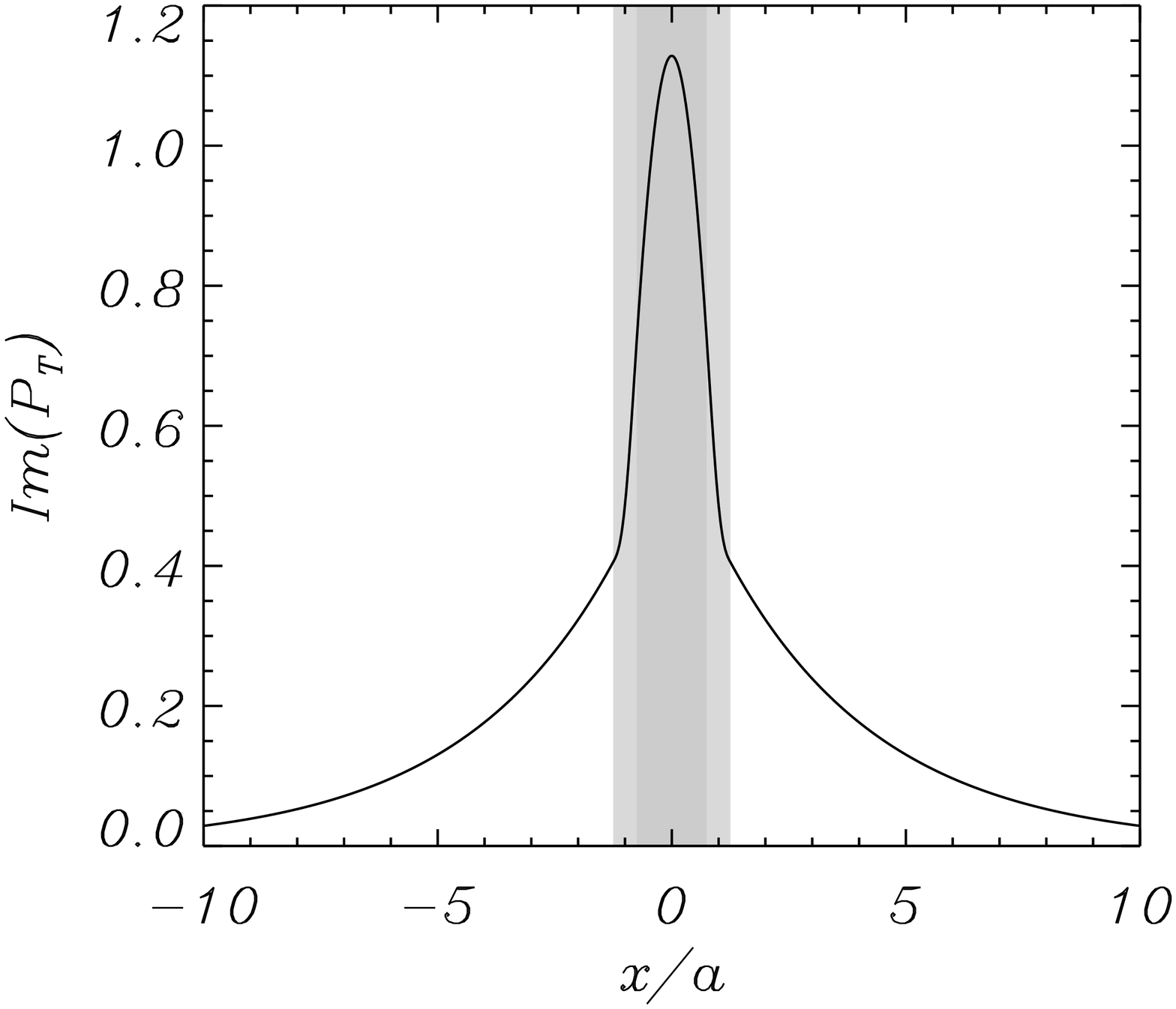}}
\hbox{
   \includegraphics[width=5.cm,angle=90]{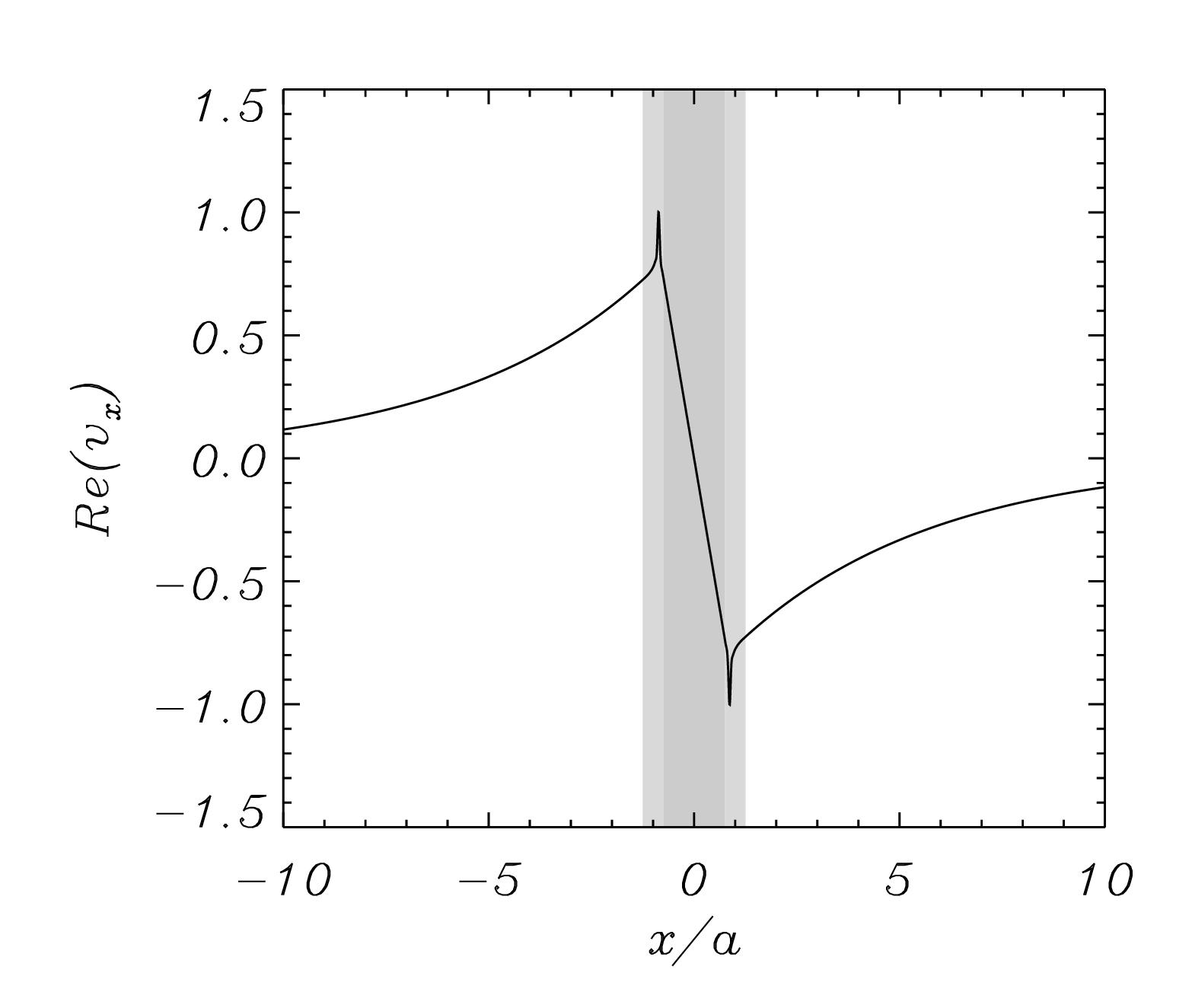}
    \includegraphics[width=5.cm,angle=90]{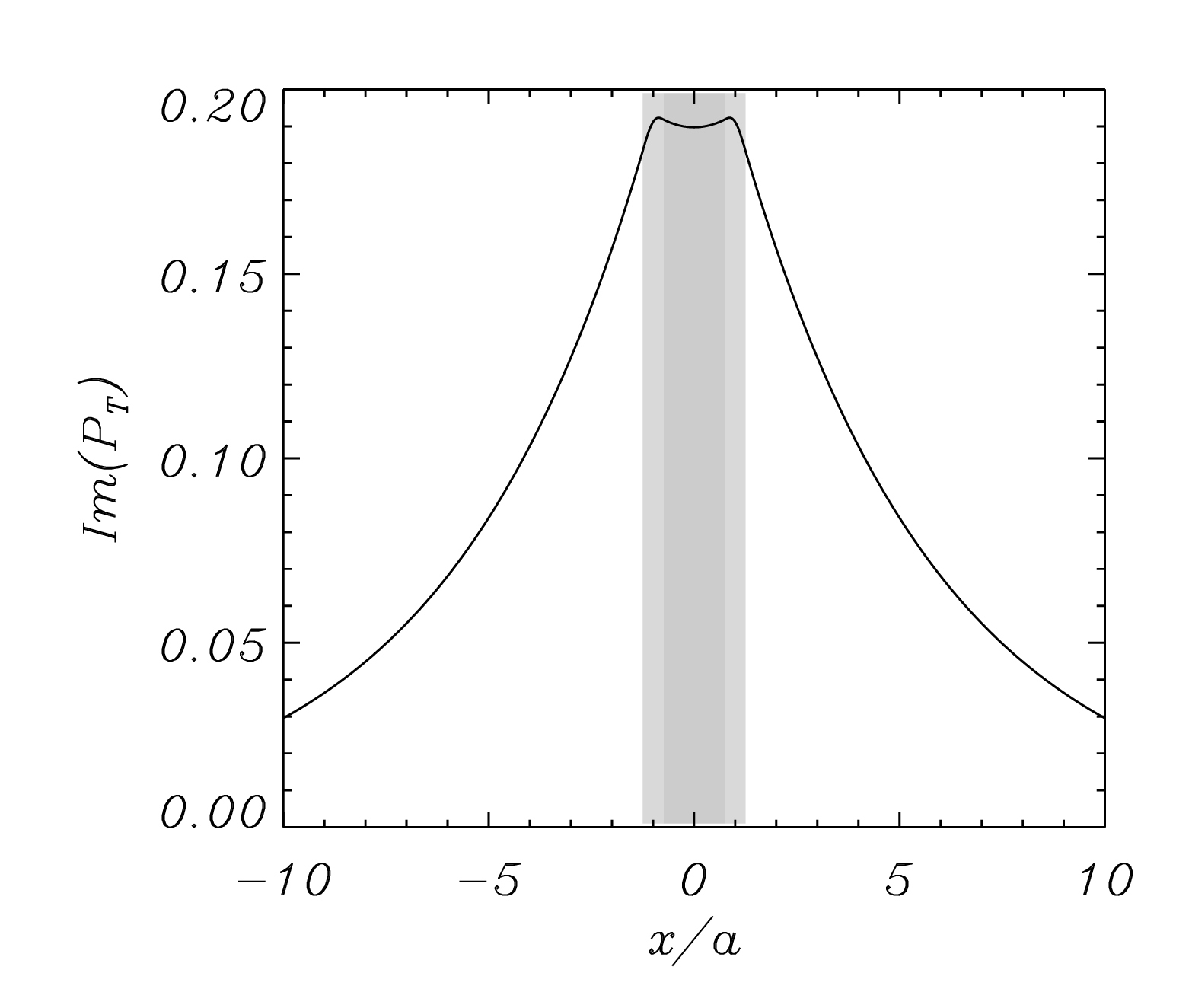}}}
 \caption{Real part of the transverse velocity component, $v_x$ and imaginary part of the perturbed 
 total pressure for the fundamental kink mode ({\em top panels}), the sausage body mode 
({\em middle panels}) and the sausage surface mode ({\em bottom panels}) for an equilibrium 
 configuration with $k_z a=\pi/50$, $\rho_i/\rho_e=10$,  $k_ya=0.5$ and non-uniform transitional 
 layers with thickness  $l/a=0.5$. A value for the magnetic Reynolds number of 
 $R_{m}/v_{Ai}a=10^{7}$ has been used in the computation of the solutions in a non-uniform grid 
with $N_x=10\ 000$ points in the range $-50$ $\le x/a \le$ $50$.
In all figures, the shaded region represents the density 
enhancement and the light-shaded regions the non-uniform layers.}
         \label{eigendamping}
\end{figure*}

The same parameter values for the equilibrium configurations are considered. Non-uniform transitional 
layers of thickness $l/a=0.5$ have been considered and we have computed the real and imaginary parts 
of the frequency for the kink mode and the sausage surface and body modes, for varying perpendicular 
wave number. The results are displayed in Figure~\ref{damping}. When we increase $k_y$, 
the damping rates of the kink mode and the sausage surface mode increase. Both damping rates are 
very similar, despite the different symmetry of the motions. The damping rate corresponding to 
the sausage body mode is roughly four to five orders of magnitude smaller than the one corresponding to 
the two surface modes and practically of the order of the used magnetic
diffusivity. The reason for the absence of resonant damping in the case of the sausage 
body solution is that its frequency, for the case of $k_z a$ small, lies 
outside the Alfven continua produced by the presence of the non-uniform 
layers  defined by the interval
$[V_{Ai}k_z,V_{Ae}k_z]$ (see
Figure~\ref{disperplots}).
In conclusion, surface-like kink and sausage oscillations are likely 
to be damped out rapidly, by resonant conversion of energy,  while
body-like sausage modes are unaffected by resonant absorption. 

Figure~\ref{eigendamping} displays example eigenfunctions obtained from the computed resonantly 
damped eigensolutions. All three modes under consideration, the sausage surface and body modes and 
the kink mode, are displayed. The spatial distribution of eigenfunctions is rather similar to the 
ones shown in Figure~\ref{eigen}, but now peaks at the non-uniform transitional layers are 
clearly visible at the perturbed transverse velocity of the kink mode and the sausage 
surface mode, while they are absent in the case of the sausage body mode. 
These peaks are an indication of resonant coupling to Alfv\'en waves at the 
non-uniform layers. Aside from these peaks, the spatial distribution of the magnetic 
pressure perturbation has extrema at the resonant layers in the case of the kink mode and 
the sausage surface mode, while it is maximum at the centre of the slab in the case
of the sausage body mode. The total pressure perturbation has no extrema at the resonant layers in the case of the 
sausage body solution. The behaviour of the eigenfunctions at the resonant layers and the fundamental conservation laws
that govern the resonant couplings were studied by Sakurai, Goossens,
and Hollweg (1991) in ideal MHD and  by Goossens, Ruderman, and Hollweg (1995) in dissipative MHD. 
These authors find that the derivative of the total pressure 
perturbation is zero at the resonant positions. This behaviour is retrieved in the pressure perturbation 
for our damped surface-like solutions, but not in the case of the
sausage body solutions, for which resonant couplings are absent.  

\section{Time-Dependent Analysis}\label{timesect}

The normal-mode analysis has provided us with information on the spatial
scales of the eigenmodes of the system and their relation to the 
frequency of the eigenmodes. The issue of the excitation of oscillations is now considered.
Observed transverse coronal-loop oscillations have been found to be mainly produced
by impulsive events, such as filament eruptions or flare-generated blast waves.
Previous studies have considered the temporal evolution of initial pulses in a magnetic slab.
For instance, Murawski and Roberts (1993a, b) studied the energy leakage associated with the 
propagation of sausage and kink waves in smoothed slabs and the temporal evolution of 
impulsively-generated linear waves, respectively. Murawski, Aschwanden, and Smith (1998) studied the 
different propagation phases (periodic, quasi-periodic, and decay) of impulsively-generated 
waves in solar coronal loops, with arbitrary plasma-$\beta$. More recently, Nakariakov, Pascoe, 
and Arber (2005) have demonstrated the possibility of remote diagnostics of the steepness of 
the transverse density profile and the density contrast in coronal loops, with impulsively-generated, 
short-period, fast magneto-acoustic wave trains.
From a theoretical point of view, it is of interest to study the distribution of the energy
of an initial perturbation among the different normal modes of the system, as well as the study 
of the conditions under which one or other type of mode of oscillation is excited in the system.
The amount of energy that is deposited in the trapped fast mode oscillation, for different 
initial excitations, has been studied by Terradas, Andries, and Goossens (2007), for a 
cylindrical magnetic tube. These authors find that the amount of energy deposited in the loop depends 
on the shape and distance of the initial perturbation. In this paper, we are interested in the 
conversion of energy from fast modes to Alfv\'en modes and the comparison of the oscillatory properties 
of the excited excited disturbances, such as the period and the damping, with the results 
of the normal mode analysis. For this reason, we show results of numerical simulations of the 
linearised MHD equations performed with initial disturbances that deposit a considerable 
amount of energy in the loop. As the normal modes studied in Section~\ref{normal} 
are characterised by their symmetry around $x=0$, it is likely that an initial disturbance with a 
given symmetry will excite one or the other types of mode. For this reason we have performed 
numerical simulations with two different types of initial disturbances, namely $v_x$ symmetric and 
antisymmetric with respect to $x=0$.

To solve Equations~(\ref{first})--(\ref{last}) numerically we have used the time-dependent version 
of the PDE2D code. Also, $\eta$ is now set to zero. The code uses a second-order 
Crank-Nicholson method with adaptive time-step control. Since we are considering a finite domain, 
reflections at the domain boundaries may affect the dynamics of the simulated loop. We have 
solved this problem by locating the edges of the numerical domain far enough from the loop. 
Given that the size of the domain is much larger than the loop thickness, we have used a non-uniform 
grid with 10\ 000 grid points in the full domain, one fifth of them located inside the loop and 
another  fifth in each inhomogeneous layer.

\begin{figure*}[!t]
 \hspace{-0.7cm}
\vbox{\hbox{
\includegraphics[width=5.2cm,angle=90]{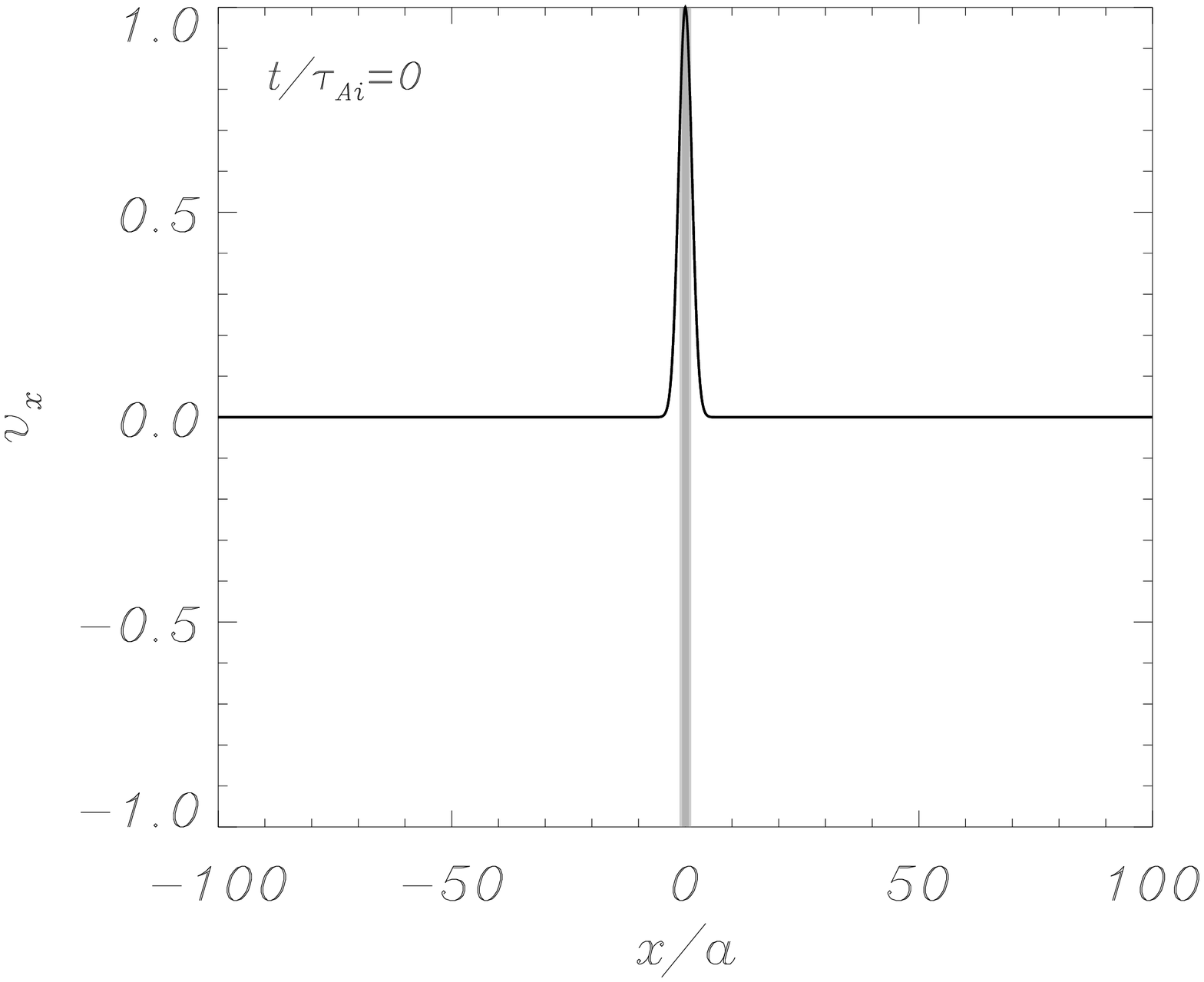}
  \includegraphics[width=5.2cm,angle=90]{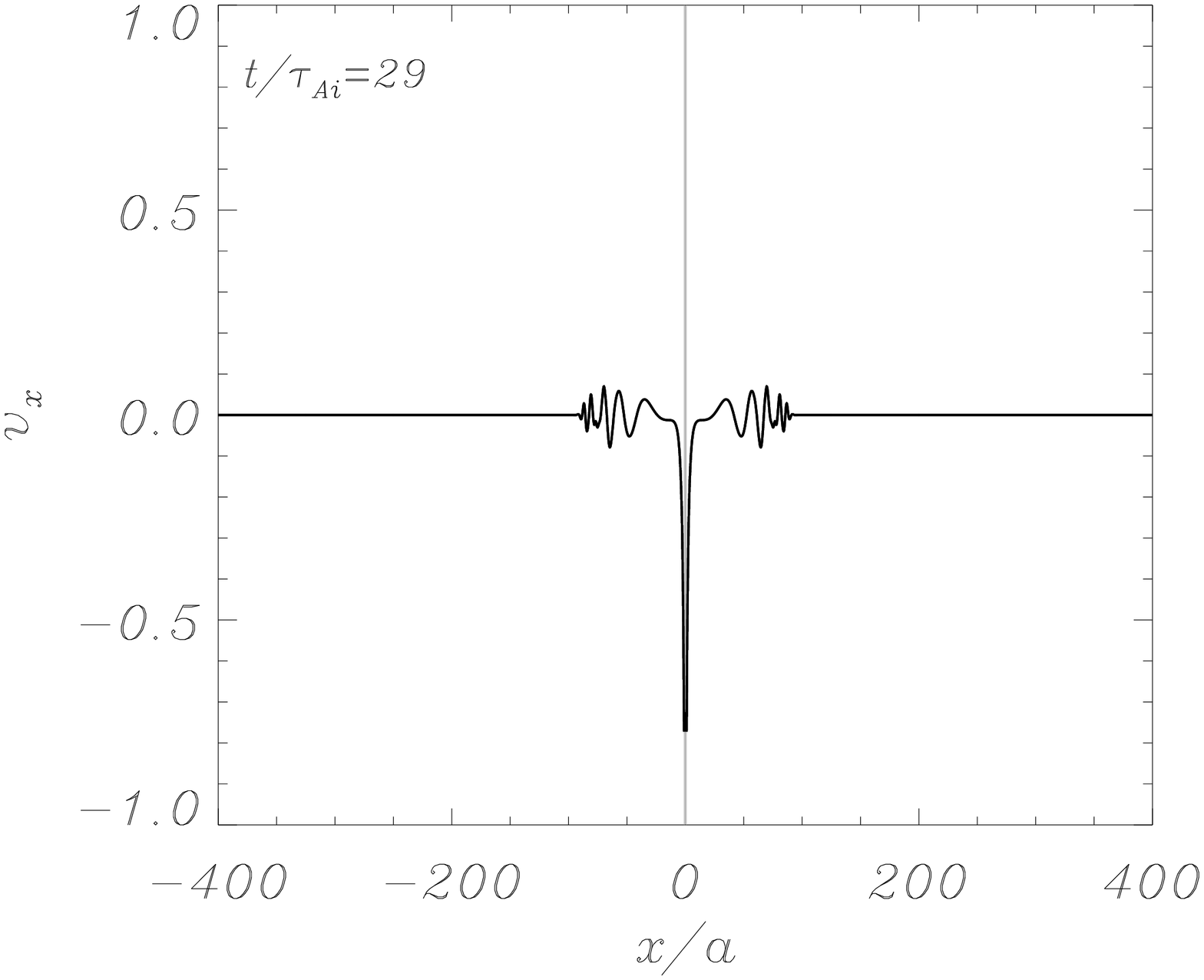} }
\hbox{
   \includegraphics[width=5.2cm,angle=90]{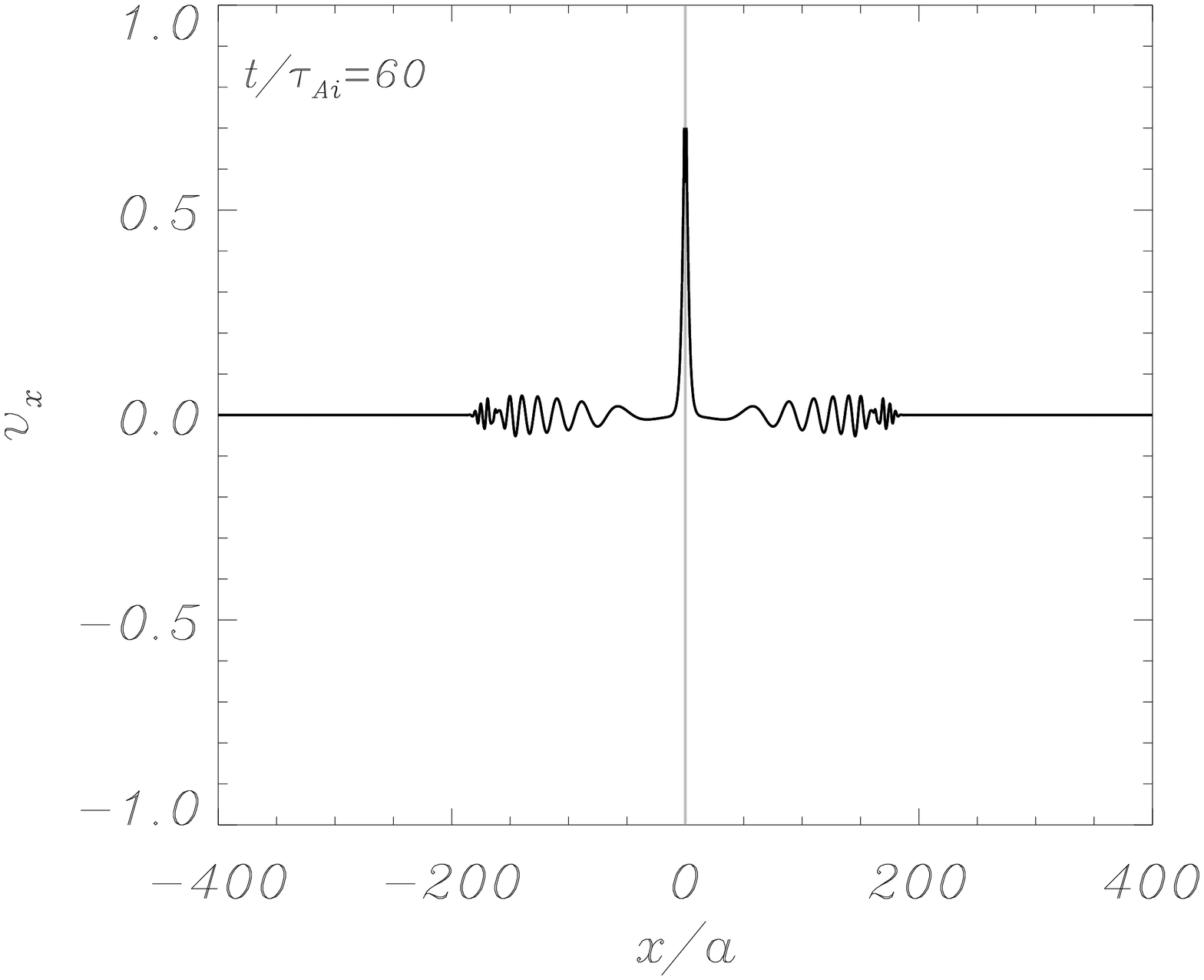}
    \includegraphics[width=5.2cm,angle=90]{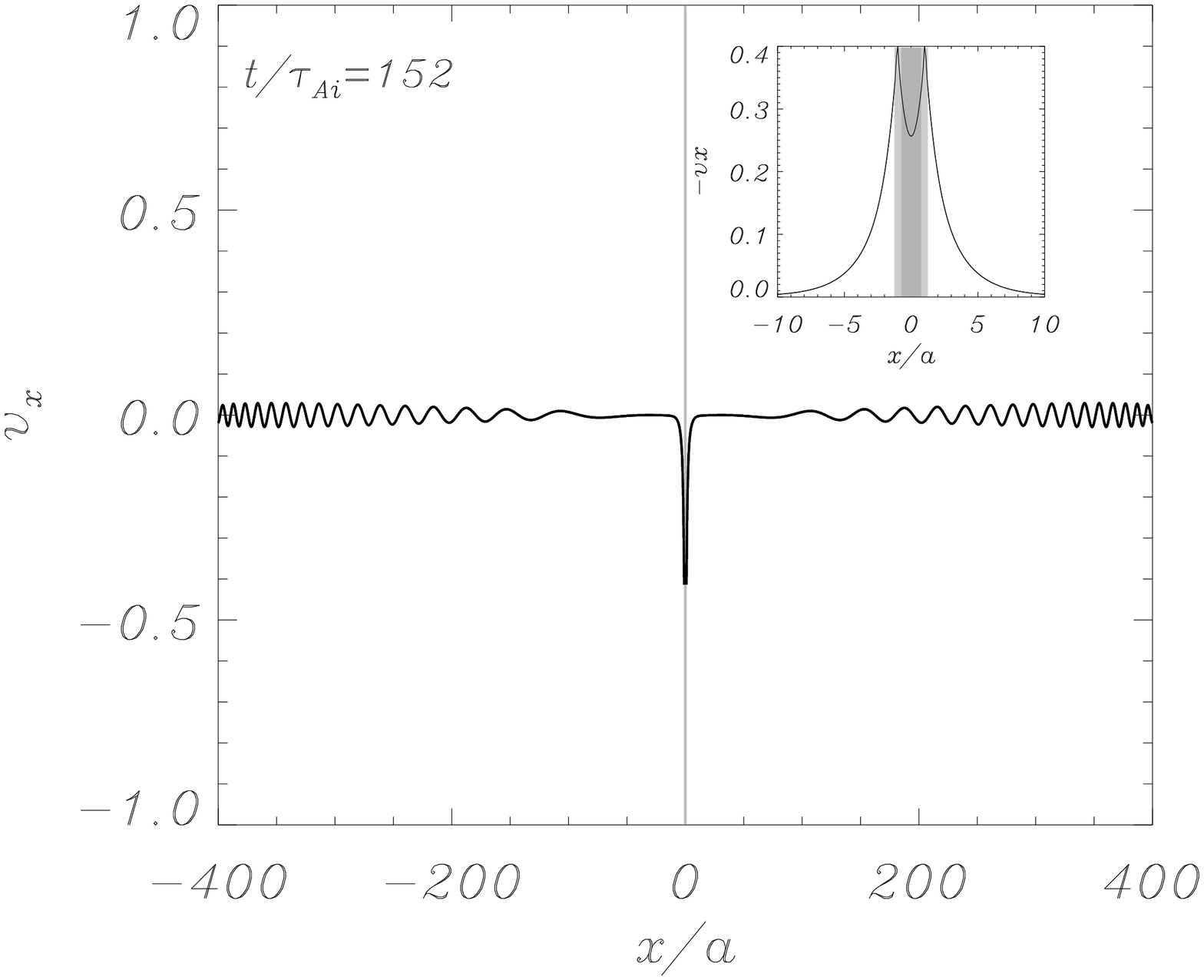}}}
\caption{Transverse velocity component, $v_x$, at different times (in units of the 
internal Alfv\'en transit time, $\tau_{Ai}=a/v_{Ai}$) for a symmetric disturbance given by 
Equation~(\ref{kinkpert}), with $x_0=0$, $w=2a$ and $a=1$. The inset plot in the lower-sight frame  
displays a detailed view of the transverse velocity component of the excited kink mode in order to 
compare it with the eigenmode computation shown in Figure~\ref{eigendamping}. The grey-shaded regions 
represent the loop.}
         \label{timeevolkink}
\end{figure*}

\begin{figure*}[!t]
 \hspace{-0.6cm}
\includegraphics[width=5.4cm,angle=90]{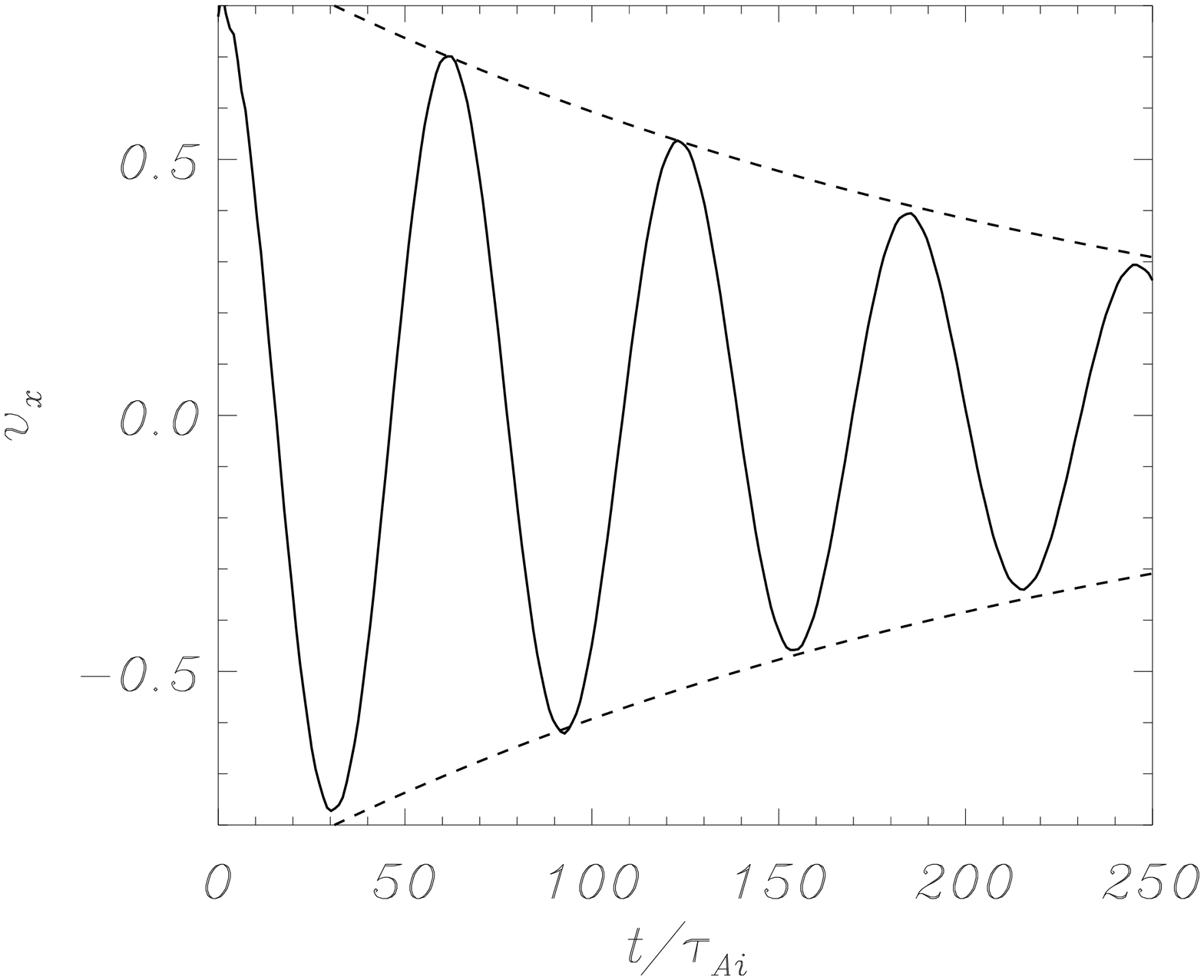}
\hspace{-0.4cm} 
\includegraphics[width=5.4cm,angle=90]{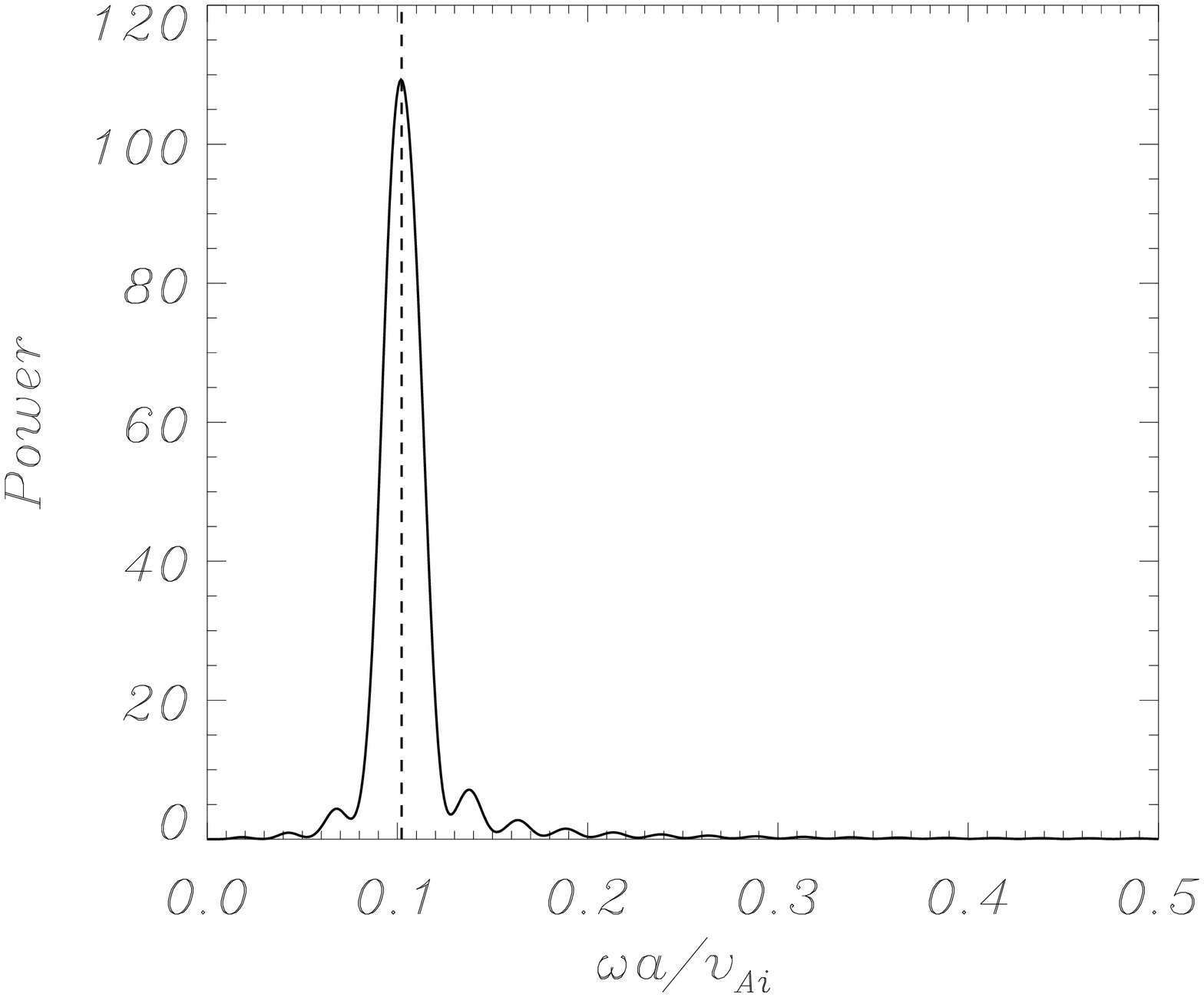}\\
\caption{{\em Left panel}: Transverse velocity [$v_x$] at the slab centre [$x=0$] as 
a function of time for the simulation shown in Figure~\ref{timeevolkink}. After a very short 
transient phase the loop reaches a stationary state, oscillating at the frequency predicted by the
normal mode analysis. The oscillation is damped due to resonant absorption. {\em Right panel}: 
Periodogram of the signal, that peaks at $\omega a/v_{Ai}=0.1018$, in good agreement with 
the normal mode result, $\omega_{R} a/v_{Ai}=0.1022$ (dashed-line). 
An exponential fit of the signal for $v_x$ gives a damping time $\tau_{d}/\tau_{Ai}=225.5$, 
that corresponds to an imaginary part of the frequency $\omega_{I} a/v_{Ai}=4.43$ $\times$ $10^{-3}$, 
in good agreement with the normal mode result, $\omega_{I}a/v_{Ai}=4.38$ $\times$ $10^{-3}$. 
Since both the peak power and the damping time agree with the normal-mode results, there 
is a clear evidence that the fundamental fast kink body mode has been excited.}
         \label{timekink}
\end{figure*}

\subsection{Symmetric Excitation}

We first consider an initial disturbance on the transverse velocity component with a Gaussian of 
the form

\begin{equation}\label{kinkpert}
v_{x}(x,t=0)=v_{x0}\left\{\exp\left[-\left(\frac{x-x_0}{w}\right)^2\right]\right\},
\end{equation}

\noindent
where $v_{x0}$ is the amplitude of the perturbation, $x_0$ the position of the Gaussian centre 
and $w$ its width at half-height. This form of excitation aims to simulate the transverse 
disturbance generated by a flare or filament eruption. Since the numerical computations apply to 
the linear regime, the particular value of $v_{x0}$ is not relevant and has been set to $v_{Ai}$.

In order to compare our results with the properties of the eigensolutions described in 
Section~\ref{normal},  we have considered a slab model of half-width $a$, density contrast 
$\rho_i/\rho_e=10$, and $k_za=\pi/50$. For the perpendicular wave number and the non-uniform 
transitional layers thickness we take values of $k_ya=0.5$ and  $l/a=0.5$. The system is then 
disturbed with a perturbation given by Equation~(\ref{kinkpert}) with $x_0=0$ and $w=2a$. 
The results of the simulation are shown in Figure~\ref{timeevolkink}, where the spatial distribution 
of the transverse velocity component is plotted at different times. The initial perturbation 
produces travelling disturbances to the left and right of its initial location and these 
disturbances exhibit some dispersion as they propagate. After a short time it is seen that 
the distribution of energy has a maximum at the centre of the system, with the amplitude 
decreasing as we move away from the edges of the slab. This indicates that the fundamental 
kink eigenmode may have been excited. The inset plot in the last shown frame supports this idea, since
the structure of $v_x$ is almost identical to that of the normal mode, shown in Figure~\ref{eigendamping}.
In Figure~\ref{timekink} (left)  
the velocity at  $x=0$ is plotted. The signal at this position clearly shows a damped oscillatory 
behaviour. This damping is due to the resonant coupling and conversion of energy to Alfv\'enic 
motions. We have performed a periodogram (Figure~\ref{timekink}, right)  and found that the 
dominant period agrees very well with the period of the fundamental kink eigenmode obtained 
in Section~\ref{normal}. The extrema of the signal are then fitted to an exponential of the form $A\ e^{-\tau_d/ t}$.
The fitted function is also shown in Figure~\ref{timekink} (left). 
From the periodogram and the exponential fit  we find that the dominant period and damping time 
agree very well with the oscillatory properties of the kink eigenfunction, which leads to the conclusion 
that the resonantly damped fundamental kink mode has been excited. 
Therefore, a symmetric excitation is likely to excite the fundamental kink mode of oscillation of 
the system.

\subsection{Antisymmetric Excitation}

\begin{figure*}[!t]
 \hspace{-0.7cm}
\vbox{\hbox{
\includegraphics[width=5.2cm,angle=90]{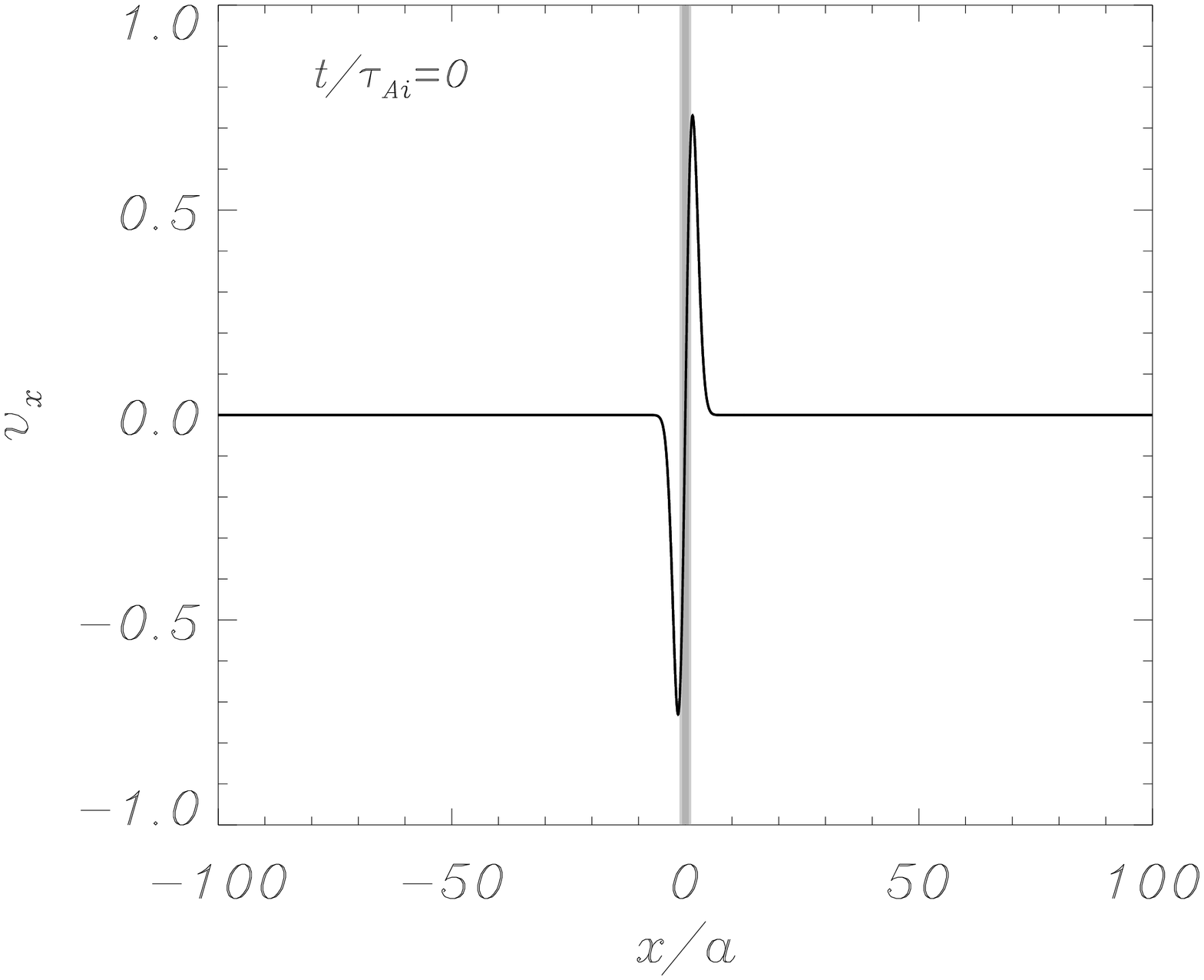}
  \includegraphics[width=5.2cm,angle=90]{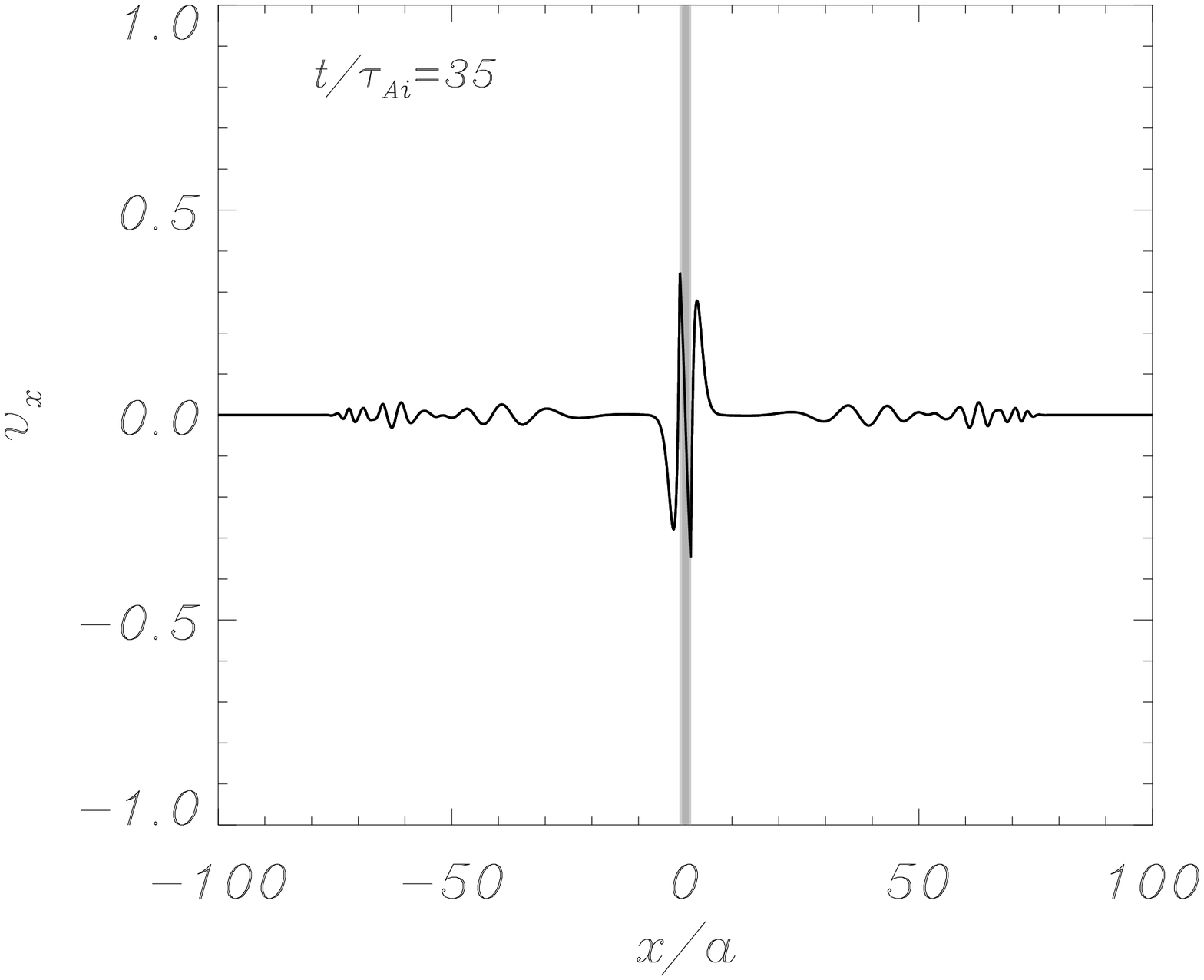} }
\hbox{
   \includegraphics[width=5.2cm,angle=90]{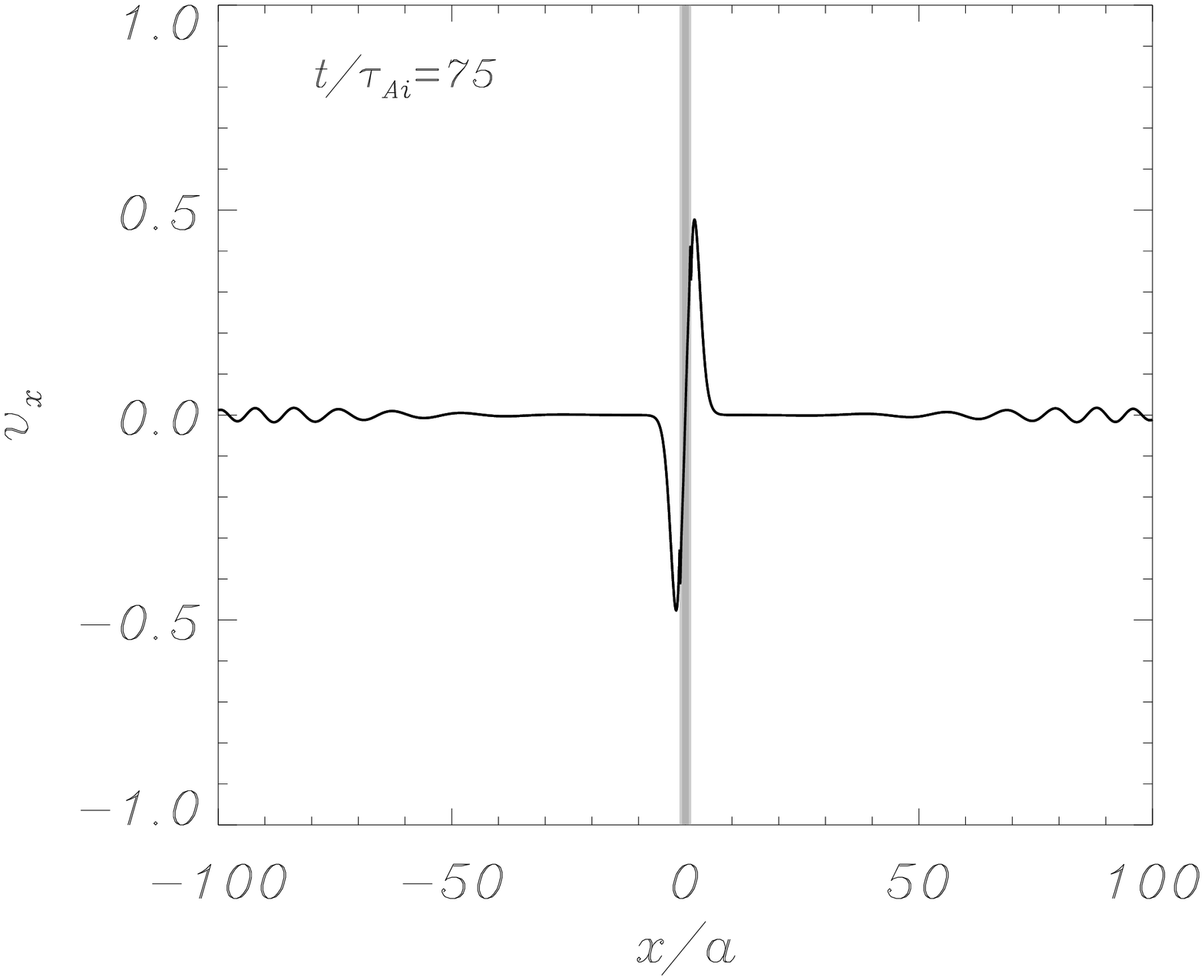}
    \includegraphics[width=5.2cm,angle=90]{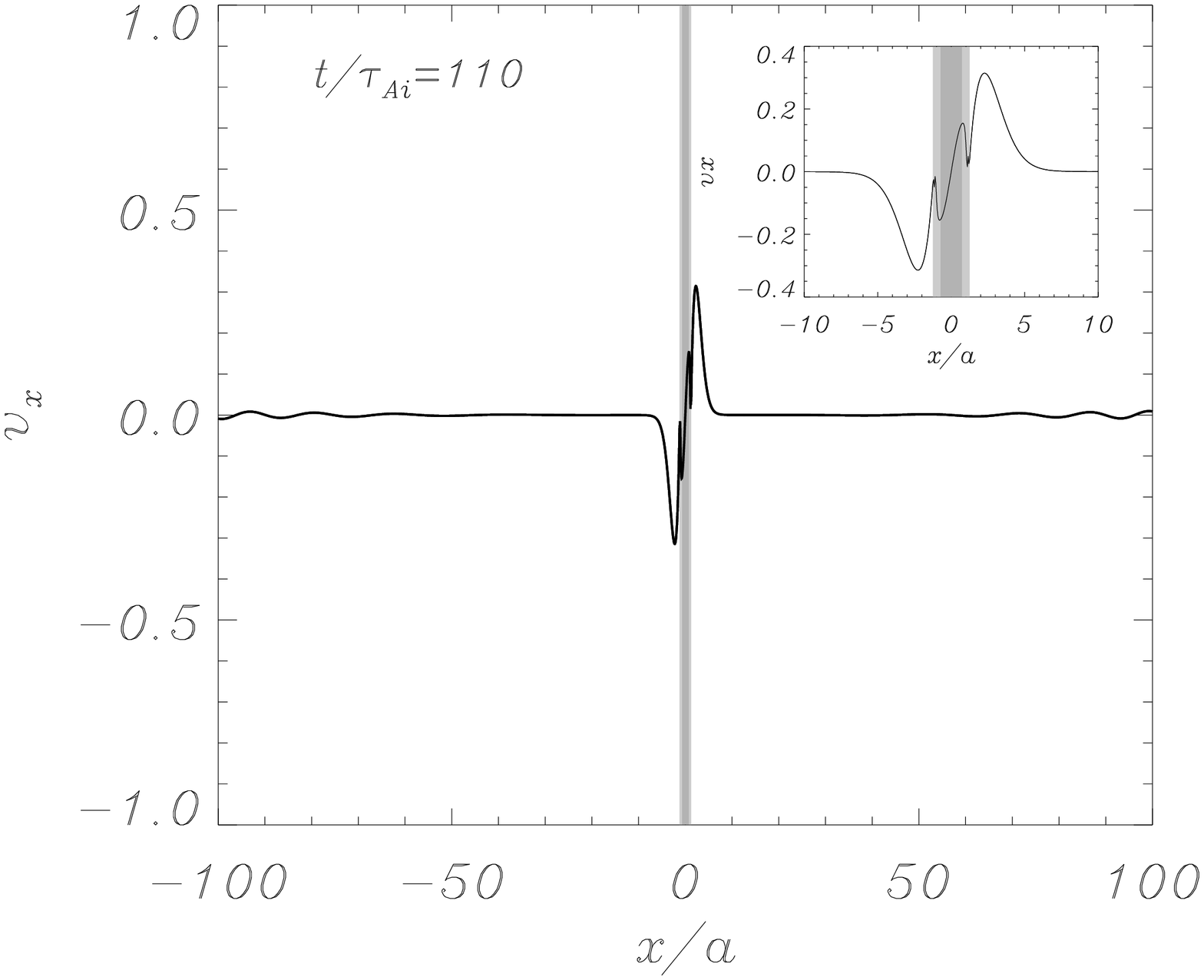}}}
\caption{Transverse velocity component [$v_x$] at different times (in units of the internal 
Alfv\'en transit time, $\tau_{Ai}=a/v_{Ai}$), for an antisymmetric disturbance given by 
Equation~(\ref{sausagepert}), with $x_0=a$, $w=2a$, and $a=1$. The inset plot in the lower-sight frame  
displays a detailed view of the transverse velocity component of the excited sausage modes in order to 
compare it with the eigenmode computations shown in Figure~\ref{eigendamping}. The grey-shaded regions 
represent the loop.}
         \label{timeevolsausage}
\end{figure*}

\begin{figure*}[!t]
 \hspace{-0.6cm}
\includegraphics[width=5.4cm,angle=90]{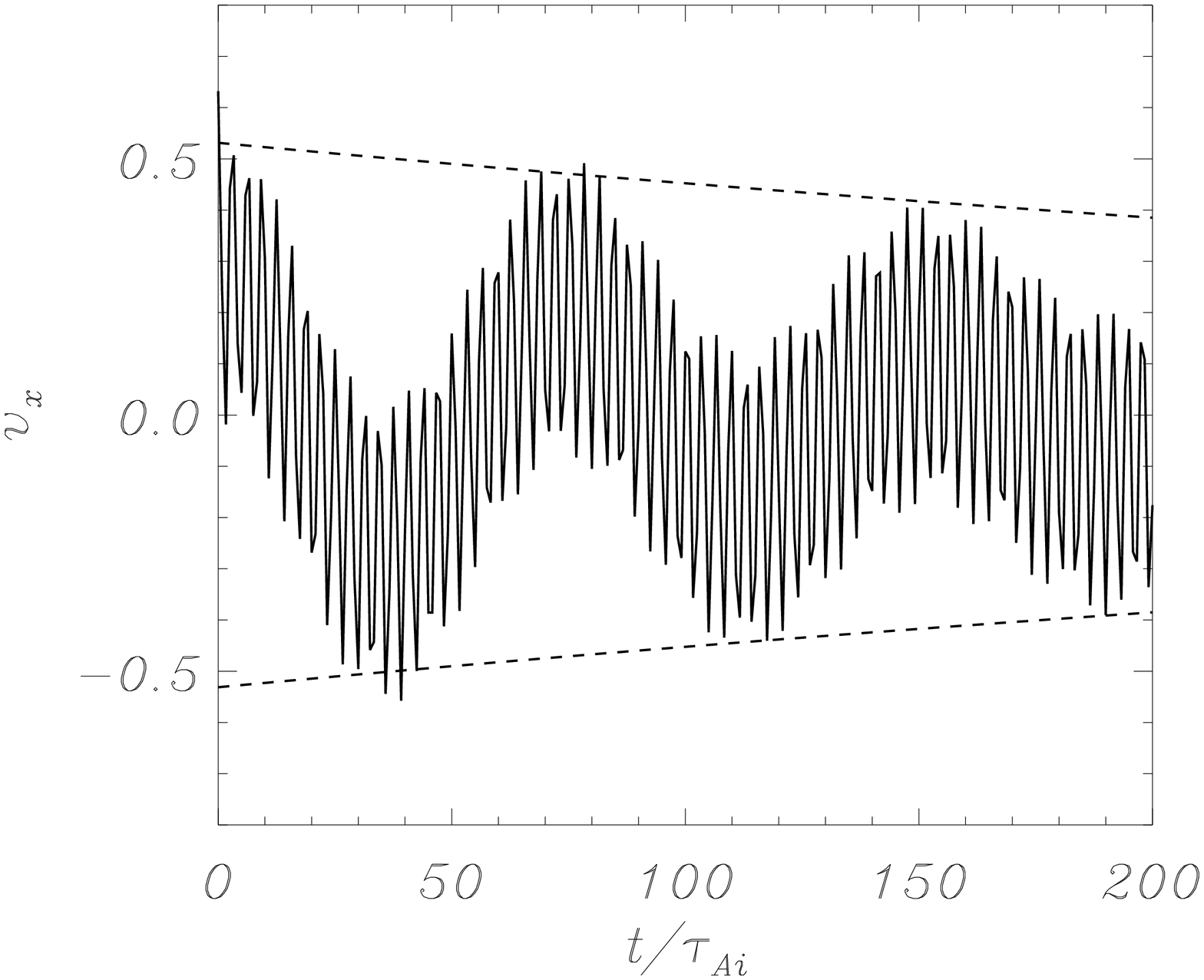}
\hspace{-0.4cm} 
\includegraphics[width=5.4cm,angle=90]{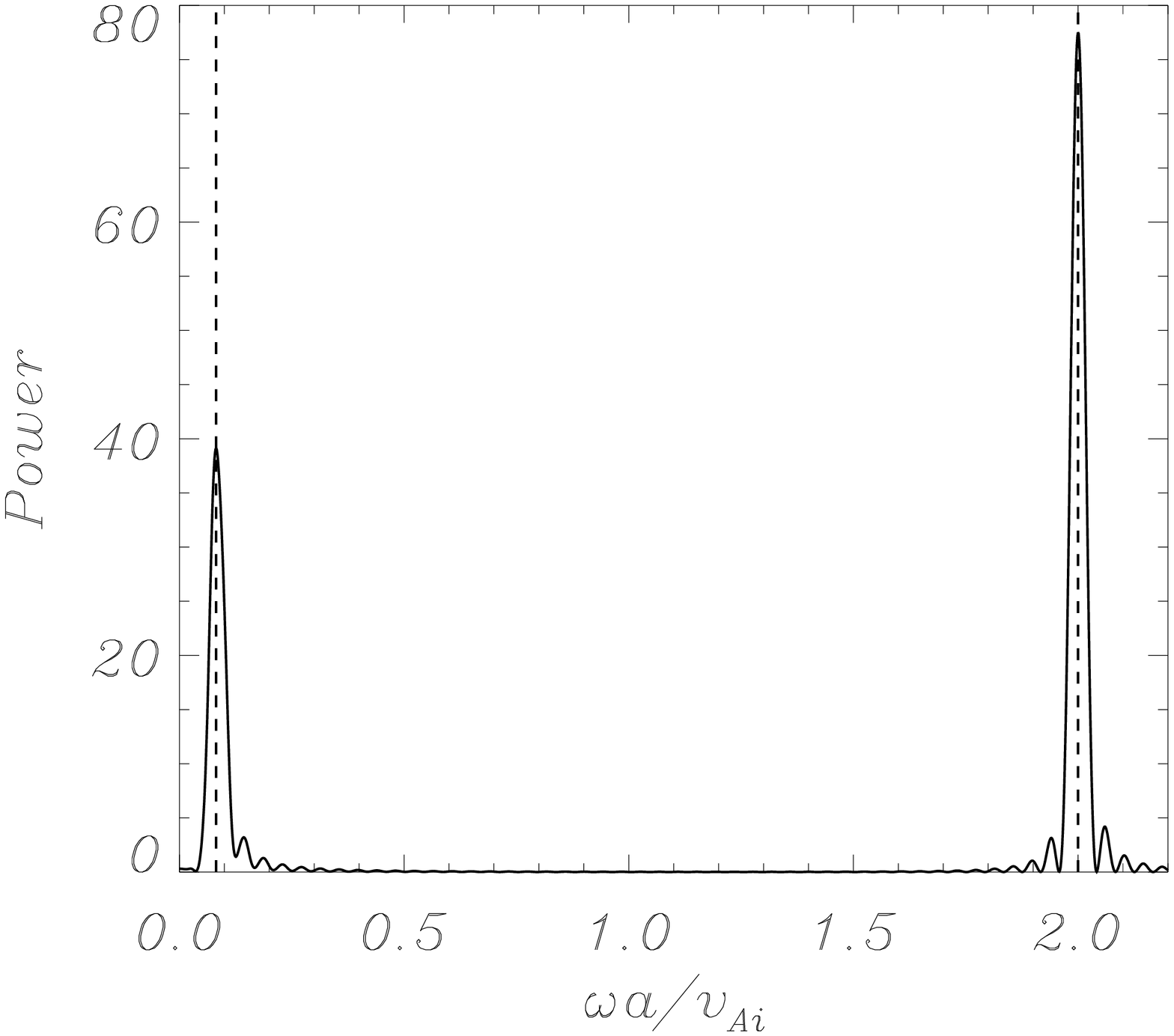}\\

 \caption{{\em Left panel}: Transverse velocity [$v_x$] at the slab edge [$x=a$] as a function 
of time for the simulation shown in Figure~\ref{timeevolsausage}. After a very short transient phase 
the loop oscillates with a linear superposition of two frequencies. The signal is damped due to resonant 
absorption. A fit of the signal for $v_x$ gives a damping time $\tau_{d}/\tau_{Ai}=621.6$, that corresponds 
to an imaginary part of the frequency $\omega_{I} a/v_{Ai}=1.6$$\times$$10^{-3}$. {\em Right panel}: 
Periodogram of the signal that peaks at $\omega_1 a/v_{Ai}=0.0817$ and  $\omega_2 a/v_{Ai}=1.9998$.  The 
dashed lines represent the frequencies obtained with the normal mode analysis; $\omega_{R} a/v_{Ai}=0.0778$ for 
the sausage surface mode and $\omega_{R} a/v_{Ai}=2.0087$ for the sausage body mode. As the damping time of 
the motion is a linear combination of the damping of the two modes the fitted damping is in between the 
imaginary parts of the frequency for the sausage surface mode ($\omega_{I1} a/v_{Ai}=1.35$$\times$$10^{-2}$) 
and the sausage body mode ($\omega_{I2} a/v_{Ai}=1.74$$\times$$10^{-6}$), for this case. Since the power 
peaks agree with the normal mode results there is a clear evidence that both the surface and body sausage 
modes have been excited.}
         \label{timesausage}
\end{figure*}

Next, we consider the excitation of the slab with an odd perturbation of the form

\begin{equation}\label{sausagepert}
v_{x}(x,t=0)=v_{x0}\left\{\exp\left[-\left(\frac{x-x_0}{w}\right)^2\right]-
 \exp\left[-\left(\frac{x+x_0}{w}\right)^2\right]\right\},
\end{equation}

\noindent
where $v_{x0}$,  $x_0$, and $w$ have the same meaning as before. The system is now disturbed with a 
perturbation given by Equation~(\ref{sausagepert}) with $x_0= a$ and $w=2a$. As the surface sausage mode 
is highly confined to the loop edges, short spatial scales are involved and these scales have to be 
introduced either through the width of the disturbance or through the value of  $k_y$. 
The first option complicates the numerical computations, so a value $k_y a=1.4$ has now been taken. 
The numerical solutions at different times (Figure~\ref{timeevolsausage}) show again that the 
induced disturbances propagate away from the slab and contain different wavelengths due to the 
dispersive nature of fast waves. After some time, the  shape of the velocity around the slab 
approaches the form of the sausage surface fast wave described in 
Section~\ref{normal}. In contrast to the kink case, the velocity distribution in the slab is now more 
complex, since more than one oscillatory mode 
seems to be present in the signal (see, for example, the inset in the last shown frame). 
The consequence of this can clearly be seen in Figure~\ref{timesausage} (left) where the 
signal at $x/a=1$ is shown. The presence of two linearly superposed oscillations is apparent. A short 
period oscillation is modulated in amplitude with a large period oscillation. The signal is again 
damped in time. The computation of the dominant periods of the signal gives the result  displayed in 
Figure~\ref{timesausage} (right). There are two peaks whose frequencies are  in perfect agreement with the 
frequencies of the sausage surface and body modes, obtained from the normal mode analysis. The power 
for the body sausage mode is larger, although this may change depending on the spatial scales involved. 
The damping time of the signal, obtained with a similar exponential fit as in the previous case, does not correspond to 
any of the two eigenmodes, since the two are present 
at the same time, but is close to the damping of the sausage surface mode. 
Therefore, we can conclude that an antisymmetric disturbance located at the edges of the slab 
induces the excitation of both types of sausage modes. The reason is that the surface and body 
sausage modes, although having very different frequencies, have very similar eigenfunctions, 
when $k_y$ is sufficiently large.

\section{Summary and Conclusions}\label{conclusions}

We have studied the oscillatory properties of surface and body MHD eigenmodes of a solar coronal loop 
including oblique propagation of perturbations. For simplicity, a Cartesian slab model has been considered 
and both the eigenvalue problem as well as the time-dependent problem have been solved.

The inclusion of oblique propagation of perturbations produces some important effects on the properties of 
eigenmodes. The resulting dispersion curves for non-zero perpendicular propagation show the existence of a 
sausage surface mode in addition to the usual fast body kink and sausage solutions, even if a 
zero-plasma-$\beta$ has been considered. This mode, not present if $k_y=0$, has the lowest frequency and 
this frequency is always below the internal cut-off frequency.  For small values of $k_y$, in the long-wave 
limit, the kink eigenmode is a body wave with the usual distribution of the velocity perturbation with a 
maximum at the centre and an evanescent behaviour outside the slab. The sausage surface mode has a velocity 
perturbation which is more confined to the edges of the slab with a short penetration depth into the coronal 
surroundings. When the value of the perpendicular wave number is increased,  a marked decrease of the 
oscillation frequency is produced in the case of the kink mode. The spatial structure of the eigenfunction 
of this mode  becomes much more confined to the slab. This leads to a change in the character of the 
kink mode which changes from body-like to surface-like. For large values of $k_y$, both surface modes 
(kink and sausage) approach the kink speed limit, that corresponds to surface waves in a magnetic interface 
in the incompressible limit or to the  kink eigenoscillations of a cylindrical flux tube in the long 
wavelength limit. As for the sausage surface mode, there is no appreciable change in its eigenfunction 
with $k_y$ and only a slight increase on the frequency is produced. The sausage body mode, which is leaky
in the long-wavelength limit, becomes trapped when, for a give value of $k_y$,  its  frequency equals
the external cut-off frequency, $\omega_{ci}$. This is due to the modification of the external 
cut-off frequency produced by the inclusion of oblique propagation. The internal cut-off frequency is also modified by
oblique propagation, but the frequency of the sausage body mode is 
always above this cut-off frequency, due to the existence of a surface mode with the same symmetry below that cut-off. 
The eigenfunctions for both sausage solutions have a marked resemblance and this has important consequences on the 
excitation of these modes. When the density is allowed to vary smoothly  between the constant internal and the constant 
external values, surface-like eigenmodes are damped by resonant conversion of energy. Both surface modes have a 
similar damping rate, while the sausage body mode is unaffected by
resonant absorption.

In the second part of the paper we have studied the time-dependent behaviour of 
our line-tied slab model for a coronal loop under different kinds of excitations.
The temporal evolution and damping by resonant conversion of wave energy of kink-  and 
sausage-type excitations has been studied for a typical coronal loop. The eigenmodes described in the 
first part of the paper can be easily excited. A symmetric disturbance located at the centre of the slab 
excites the fundamental symmetric kink mode. Depending on the value of $k_y$, the excited mode would have  
body or surface wave properties. The period of the signal 
and its damping time agree with the result obtained with the normal mode analysis. The velocity 
distribution of the stationary phase coincides with the spatial distribution of the eigenfunction. 
Interestingly, an antisymmetric initial disturbance with peaks at the edges of the slab excites both the 
surface and the body sausage modes and a linear superposition of perturbations is obtained, with the sausage 
body oscillation amplitude being modulated by the presence of  the sausage surface mode. This can be explained 
in terms of the involved eigenfunctions since  oblique propagation confines the spatial distribution of the 
body sausage mode and makes it very similar to that of the sausage surface mode.

In this work, a magnetic slab has been used to model a coronal loop. Some of the results found in slab 
geometry cannot be translated to cylindrical geometry. For instance, the sausage surface mode described in 
this paper is not present in cylindrical geometry. Also, the sausage mode of oscillation of a cylinder 
corresponds to an azimuthal wave number $m=0$, thus this mode cannot resonantly couple to Alfv\'en modes.
Slabs are known to be poor wave-guides, when compared to cylinders. However, our results indicate that the inclusion
of oblique propagation produces  a sharper drop-off rate of the kink eigenfunction in the external medium and
a frequency that is a good approximation to the kink-mode frequency in a slender flux tube.

%%%%%%%%%%%%%%%%%%%%%%%%%%%%%%%%%%%%%%%%%%%%%%%%%%%%%%%%%%%%%%%%%%%%%%%%%%%
\begin{acks}
We thank the referee, Dr.\ Erwin Verwichte, for constructive and valuable comments that 
have benefited this paper. The authors acknowledge the Spanish Ministerio de 
Educaci\'on y Ciencia for the funding 
provided under project AYA2006-07637 and the Conselleria d'Economia, Hisenda i 
Innovaci\'o of the Government of the Balearic Islands for the funding provided under 
grants PRIB-2004-10145 and PCTIB2005GC3-03. J. Terradas acknowledges the Spanish Ministerio de 
Educaci\'on y Ciencia for the funding provided under a Juan de la Cierva fellowship.
\end{acks}

\end{article} 

\begin{thebibliography}{}
\bibitem[\protect\citeauthoryear{Andries et al.}{2005}]{AAG05}
Andries, J., Arregui, I., Goossens, M.: 2005,  {\it Astrophys.
J.}  {\bf 624},
  L57.

\bibitem[\protect\citeauthoryear{Andries et al.}{2005}]{Andries05}
Andries, J., Goossens, M., Hollweg, J.V., Arregui, I., Van
  Doorsselaere, T.: 2005, {\it Astron. Astrophys.}  {\bf 430}, 1109.

\bibitem[\protect\citeauthoryear{Arregui et al.}{2005}]{Arregui05}
Arregui, I., Van Doorsselaere, T., Andries, J., Goossens, M., 
  Kimpe, D.: 2005, {\it Astron. Astrophys.}  {\bf 441}, 361.


\bibitem[\protect\citeauthoryear{Arregui et al.}{2007}]{Arregui07b}
Arregui, I., Andries, J., Van Doorsselaere, T., Goossens, M., 
  Poedts, S.: 2007b, {\it Astron. Astrophys.}  {\bf 463}, 333.


\bibitem[\protect\citeauthoryear{Aschwanden}{2006}]{Aschwanden06}
Aschwanden, M.J.: 2006, {\it Roy. Soc. London Phil. Trans.
  Ser. A}  {\bf 364}, 417.

\bibitem[\protect\citeauthoryear{Aschwanden et al.}{2002}]{Aschwanden02}
Aschwanden, M.J., De Pontieu, B., Schrijver, C.J.,  Title, A.M.:
  2002, {\it Solar Phys.}  {\bf 206}, 99.

\bibitem[\protect\citeauthoryear{Aschwanden et al.}{1999}]{Aschwanden99}
Aschwanden, M.J., Fletcher, L., Schrijver, C.J., Alexander, D.:
  1999,  {\it Astrophys. J.}  {\bf 520}, 880.
\bibitem[\protect\citeauthoryear{Aschwanden et al.}{2003}]{Aschwanden03}
Aschwanden, M.J., Nightingale, R.W., Andries, J., Goossens, M., Van Doorsselaere, T.:
  2003,  {\it Astrophys. J.}  {\bf 598}, 1375.

\bibitem[\protect\citeauthoryear{Brady and Arber}{2005}]{BA05}
Brady, C.S.,  Arber, T.D.: 2005, {\it Astron. Astrophys}.  {\bf
438}, 733.



\bibitem[\protect\citeauthoryear{Diaz}{2006}]{toni06a}
D\'{\i}az, A.J.: 2006,  {\it Astron. Astrophys.}  {\bf 456}, 737.


\bibitem[\protect\citeauthoryear{Diaz et al.}{2003}]{toni03}
D\'{\i}az, A.J., Oliver, R., Ballester, J.L.: 2003,  {\it Astron.
Astrophys.}  {\bf 402}, 781.

\bibitem[\protect\citeauthoryear{Diaz et al.}{2006}]{toni06b}
D\'{\i}az, A.J., Zaqarashvili, T.V., Roberts, B.: 2006,  {\it Astron.
Astrophys.}  {\bf 455}, 709.


\bibitem[\protect\citeauthoryear{Dymova and Ruderman}{2006}]{DR06}
Dymova, M.V., Ruderman, M.S.: 2006, {\it Astron. Astrophys.} 
{\bf 457}, 1059.

\bibitem[\protect\citeauthoryear{Edwin \& Roberts}{1982}]{ER82}
Edwin, P.M., Roberts, B.: 1982, {\it Solar Phys.}  {\bf 79}, 239.


\bibitem[\protect\citeauthoryear{Edwin \& Roberts}{1983}]{ER83}
Edwin, P.M., Roberts, B.: 1983,  {\it Solar Phys.}  {\bf 88}, 179.

\bibitem[\protect\citeauthoryear{Edwin \& Roberts}{1988}]{ER88}
Edwin, P.M., Roberts, B.: 1988,  {\it Astron. Astrophys.} {\bf
192}, 343.


\bibitem[\protect\citeauthoryear{Goossens}{1991}]{Goossens91}
Goossens, M.: 1991 In Ulmschneider, P., Priest E.R., Rosner, R.
(eds.), {\it Mechanisms of Chromospheric and Coronal Heating},
Springer-Verlag, Berlin, 480.

\bibitem[\protect\citeauthoryear{Goossens et al.}{2006}]{GAA06}
Goossens, M., Andries, J.,  Arregui, I.: 2006, {\it Roy. Soc. London
  Phil. Trans. Ser. A}  {\bf 364},  433.

\bibitem[\protect\citeauthoryear{Goossens et al.}{2002}]{GAA02}
Goossens, M., Andries, J.,  Aschwanden, M.J..: 2002, {\it Astron. Astrophys.}
{\bf 394},  L39.

\bibitem[\protect\citeauthoryear{Goossens et al.}{1995}]{GRH95}
Goossens, M., Ruderman, M.S.,  Hollweg, J.V.: 1995, 
{\it Solar Phys.}  {\bf 157},  75.


%\bibitem[\protect\citeauthoryear{Goossens et al.}{2002}]{GAA02}
%Goossens, M., Andries, J., \& Aschwanden, M.~J.: 2002, {\it Astron. Astrophys.}, {\bf 394}, L39.


\bibitem[\protect\citeauthoryear{Hollweg \& Yang}{1988}]{HY88}
Hollweg, J.V., Yang, G.: 1988, {\it J. Geophys. Res.}  {\bf 93}, 5423.

\bibitem[\protect\citeauthoryear{Hollweg}{1990}]{Holl90a}
Hollweg, J.V.: 1990a, {\it Comp. Phys. Reports} {\bf 12}, 205.

\bibitem[\protect\citeauthoryear{Hollweg}{1990}]{Holl90b}
Hollweg, J.V.: 1990b, In  Russel, C.T.,  Priest, E.R.,  Lee, L.C.
(eds.), {\it Physics of Magnetic Flux Ropes},  (Washington, AGU), Geophys.
Mono.  58, 123.

\bibitem[\protect\citeauthoryear{Hollweg}{1991}]{Holl91}
Hollweg, J.V.: 1991 In Ulmschneider, P., Priest E.R., Rosner, R.
(eds.), {\it Mechanisms of Chromospheric and Coronal Heating},
Springer-Verlag, Berlin, 423.

\bibitem[\protect\citeauthoryear{Jain \& Roberts}{1994}]{JR94}
Jain, R., Roberts, B.: 1994, {\it Astron. Astrophys.} {\bf 286},
243.

\bibitem[\protect\citeauthoryear{Lee \& Roberts}{1986}]{LR86}
Lee, M.A., Roberts, B.: 1986,  {\it Astrophys. J.}  {\bf 301}, 430.

\bibitem[\protect\citeauthoryear{McEwan et al.}{2006}]{McE06}
McEwan, M.P., Donnelly, G.R., D\'{\i}az, A.J., Roberts, B.: 2006,
{\it Astron. Astrophys.}  {\bf 460}, 893.

\bibitem[\protect\citeauthoryear{Miles \& Roberts}{1989}]{MR89}
Miles, A.J., Roberts, B.: 1989,  {\it Solar Phys.} {\bf 119}, 257.


\bibitem[\protect\citeauthoryear{Murawski \& Roberts}{1993a}]{MR93a}
Murawski, K., Roberts, B.: 1993a, {\it Solar Phys.}  {\bf 143}, 89.


\bibitem[\protect\citeauthoryear{Murawski \& Roberts}{1993b}]{MR93b}
Murawski, K., Roberts, B.: 1993b, {\it Solar Phys.} {\bf 144}, 101.


\bibitem[\protect\citeauthoryear{Murawski et al.}{1998}]{MR98}
Murawski, K., Aschwanden, M.J., Smith, J.M.: 1998, {\it Solar
Phys.} {\bf 179}, 313.


\bibitem[\protect\citeauthoryear{Nakariakov \& Ofman}{2001}]{Nakariakov01}
Nakariakov, V.M., Ofman, L.: 2001, {\it Astron. Astrophys.} {\bf
372}, L53.

\bibitem[\protect\citeauthoryear{Nakariakov \& Roberts}{1995}]{Nakariakov95}
Nakariakov, V.M., Roberts, B.: 1995,  {\it Solar Phys.}   {\bf
159}, 399.

\bibitem[\protect\citeauthoryear{Nakariakov \& Verwichte}{2005}]{Nakariakov05}
Nakariakov, V.M., Verwichte, E.: 2005, {\it Living Rev. Solar
Phys.}  {\bf 2}, 3. {\sf http://solarphysics.livingreviews.org}

\bibitem[\protect\citeauthoryear{Nakariakov, Pascoe, and Arber}{2005}]{Nakariakov05b}
Nakariakov, V.M., Pascoe, D.J.,  Arber, T.D..: 2005, 
{\it Space Sci Rev.} {\bf 121}, 115.


\bibitem[\protect\citeauthoryear{Nakariakov et al.}{1999}]{Nakariakov99}
Nakariakov, V.M., Ofman, L., DeLuca, E.E., Roberts, B.,  Davila,
  J.M.: 1999, {\it Science} {\bf 285}, 862.


\bibitem[\protect\citeauthoryear{Poedts \& Kerner}{1991}]{POKE91}
Poedts, S., Kerner, W.: 1991, {\it Phys. Rev. Lett.} {\bf 66}, 2871.

\bibitem[\protect\citeauthoryear{Roberts}{1981}]{Roberts81a}
Roberts, B.: 1981a, {\it Solar Phys.} {\bf 69}, 27.

\bibitem[\protect\citeauthoryear{Roberts}{1981}]{Roberts81b}
Roberts, B.: 1981b, {\it Solar Phys.} {\bf 69}, 39.

\bibitem[\protect\citeauthoryear{Roberts}{1983}]{Roberts83}
Roberts, B.: 1983,  {\it Solar Phys.} {\bf 87}, 77.


\bibitem[\protect\citeauthoryear{Roberts}{1991}]{Roberts91}
Roberts, B.: 1991, In Ulmschneider, P., Priest E.R., Rosner, R.
(eds.), {\it Mechanisms of Chromospheric and Coronal Heating},
Springer-Verlag, Berlin, 494.

\bibitem[\protect\citeauthoryear{Roberts et al.}{1984}]{REB84}
Roberts, B., Edwin, P.M.,  Benz, A.O.: 1984,   {\it Astrophys.
J.}  {\bf 279}, 857.

\bibitem[\protect\citeauthoryear{Ruderman}{2003}]{Ruderman03}
Ruderman, M.S.: 2003,  {\it Astron. Astrophys.}  {\bf 409}, 287.


\bibitem[\protect\citeauthoryear{Ruderman \& Roberts}{2002}]{RR02}
Ruderman, M.S. and Roberts, B.: 2002,  {\it Astrophys. J.}  {\bf
577}, 475.

\bibitem[\protect\citeauthoryear{Ryutova}{1990}]{Ryutova90}
Ryutova, M.~P.: 1990, In  Stenflo, J.O. (ed.),
{\it Solar Photosphere: Structure, Convection and Magnetic Fields}, 
( Reidel, Dordrecht), IAU Symp. 138, 229.

\bibitem[\protect\citeauthoryear{Sakurai et al.}{1991}]{SGH91}
Sakurai, T., Goossens, M., Hollweg, J.V.: 1991,  {\it Solar
Phys.} {\bf 133},
  227.


\bibitem[\protect\citeauthoryear{Sewell}{2005}]{Sewell05}
Sewell, G.: 2005,  {\it The Numerical Solution of Ordinary and Partial
Differential Equations}, John Wiley \& Sons, Inc.,
Hoboken, New Jersey.


\bibitem[\protect\citeauthoryear{Schrijver et al.}{2002}]{SAT02}
Schrijver, C.J., Aschwanden, M.J.,  Title, A.M.: 2002,  {\it
Solar Phys.} {\bf 206},
  69.
\bibitem[\protect\citeauthoryear{Spruit}{1981}]{Spruit81}
Spruit, H.C.: 1981, In Jordan, S. (ed.), {\it The Sun as a Star}, 
SP-450, NASA Washington, 385.

\bibitem[\protect\citeauthoryear{Terradas, et al.}{2005}]{TOB05}
Terradas, J., Oliver, R., Ballester, J.L.: 2005,  {\it Astron.
Astrophys.}  {\bf 441}, 371.



\bibitem[\protect\citeauthoryear{Terradas, et al.}{2006}]{TOB06}
Terradas, J., Oliver, R., Ballester, J.L.: 2006,  {\it
Astrophys. J.}  {\bf 650}, L91.

\bibitem[\protect\citeauthoryear{Terradas, et al.}{2007}]{TAG07}
Terradas, J., Andries, J., Goossens, M.: 2007,  {\it Astron.
Astrophys.}, {\bf 469}, 1135.



\bibitem[\protect\citeauthoryear{}{}]{}
Uchida, Y.: 1970,  {\it Pub. Astron. Soc. Japan} {\bf 22}, 341.

\bibitem[\protect\citeauthoryear{Van Doorsselaere et al.}{2004}]{tom04b}
Van Doorsselaere, T., Andries, J., Poedts, S., Goossens, M.: 2004,
 {\it Astrophys. J.}  {\bf 606}, 1223.


\bibitem[\protect\citeauthoryear{Verwichte et al.}{2006a}]{verwichte06a}
Verwichte, E., Foullon, C., Nakariakov, V.M.: 2006a,  {\it
Astron. Astrophys.}  {\bf 446}, 1139.

\bibitem[\protect\citeauthoryear{Verwichte et al.}{2006b}]{verwichte06b}
Verwichte, E., Foullon, C.,  Nakariakov, V.M.: 2006b,  {\it
Astron. Astrophys.} {\bf 449}, 769.


\bibitem[\protect\citeauthoryear{Verwichte et al.}{2006c}]{verwichte06c}
Verwichte, E., Foullon, C.,  Nakariakov, V.M.: 2006c,  {\it
Astron. Astrophys.} {\bf 452}, 615.


\bibitem[\protect\citeauthoryear{Zhelyazkov et al.}{1996}]{zhelyazkov96}
Zhelyazkov, I., Murawski, K.,  Goossens, M.: 1996,  {\it Solar
Phys.} {\bf 165}, 99.

\end{thebibliography}
\end{document}